\documentclass[letter,oneside,9pt,twocolumn,article]{aiaa}
\usepackage{graphicx}
\usepackage{rotating}
\usepackage{bm}
\usepackage{flushend}
\usepackage[bf,BF,small]{subfigure}
\usepackage{adjustbox}
\usepackage{tabularx}
\usepackage{multirow}
\usepackage{multicol}
\usepackage{indentfirst} 
\usepackage[hidelinks]{hyperref}
\usepackage{float}
\usepackage{xcolor}
\usepackage[round]{natbib}
\setcitestyle{authoryear,open={(},close={)}} 

\usepackage{mwe}
\usepackage[markcase=noupper]{scrlayer-scrpage}
\ohead*{\pagemark}
\ihead*{Physics of Fluids 36, 086113 (2024); doi: 10.1063/5.0219552}
\ifoot*{\rm ~~}
\ofoot*{\rm ~~}
\newcommand{\alb}{\vspace{0.1cm}\\} 
\newcommand{\mfd}{\displaystyle}

\newcommand{\ns}{{n_{\rm s}}}

\renewcommand{\fontsizetable}{\footnotesize\scalefont{0.99}}

\renewcommand{\vec}[1]{\bm{#1}}
\author{ Bernard Parent\thanks{Associate Professor, Aerospace and Mechanical Engineering, University of Arizona, Tucson, AZ 85721, USA, bparent@arizona.edu, author to whom correspondence should be addressed.}
~~and~ 
Felipe Martin Rodriguez Fuentes \thanks{Graduate Student, Aerospace and Mechanical Engineering, University of Arizona, Tucson, AZ 85721, USA, fmrodriguez@arizona.edu} 
\\
          {\it University of Arizona, Tucson, AZ 85721}\\
    }

\title{
Progress in Electron Energy Modeling\\
for Plasma Flows and Discharges
}

\abstract{
A novel formulation of the electron energy relaxation terms is presented here, which is applicable to plasma flows and discharges wherein the electron temperature could be higher or lower than the gas temperature. It is demonstrated that the electron energy losses due to inelastic collisions can be expressed as a function of only two species-dependent parameters: the reduced electric field and the reduced electron mobility. This formulation is advantageous over previous ones, being simpler to implement and more accurate when experimental data of the reduced electric field and reduced mobility are available. Curve fits to empirical data of these two properties are outlined here for all important air molecular species. The approach accounts for all inelastic electron energy relaxation processes without needing individual cross-sections or rates, reducing potential errors associated with independently handling each process. Several test cases are presented to validate the proposed electron energy source terms including re-entry plasma flows for which the electron temperature is less than the gas temperature, as well as discharges in which the electron temperature reaches values in excess of 30~eV.  In all cases, the agreement with experimental data is observed to be very good to excellent, significantly surpassing   prior electron energy models for plasma flows.
}

\nomenclature{

\begin{nomenclaturelist}{}
\item[$a$]{local speed of sound (${\rm m/s}$)}
\end{nomenclaturelist}

 \begin{nomenclaturelist}{Subscripts}
  \item[$t$] turbulent contribution
   \end{nomenclaturelist}
  
\begin{nomenclaturelist}{Superscripts}
 \item[$\star$] sum of molecular and turbulent contributions
  \end{nomenclaturelist}
}
\begin{document}
\maketitle

\section{Introduction}
\dropword At flight Mach numbers in excess of 10, significant quantities of electrons and ions are created near the nose of hypersonic vehicles through the associative ionization of nitric oxide. Due to the relatively slow electron-ion recombination process, a plasma surrounds the entire vehicle. The presence of plasma poses problems, as it can lead to radio communication blackout [\cite{jsr:2010:kim,ieee:1971:rybak, aiaaconf:2007:hartunian, aiaaconf:2011:davis}] and to a large ionized wake  detectable with radar [\cite{ieee:2019:wang, aiaa:1964:lees}]. A plasma surrounding the vehicle can also be advantageous by absorbing some incoming electromagnetic waves [\cite{book:2016:singh, spie:2020:xu, gmrl:1963:musal}].

The electron temperature plays a critical role in determining plasma density in three different ways. First, it can affect the production of electrons through Townsend ionization because the latter reaction rate depends on electron temperature. Two, it can affect the destruction of electrons through electron-ion two-body and three-body recombination because both reactions are function of electron temperature. Third, it can affect plasma density through electron loss to the surfaces. Indeed, as shown in \cite{pof:2022:parent}, electron loss to the surface of the vehicle due to catalytic effects is not negligible in hypersonic flows and can even dominate over electron loss due to chemical reactions within the plasma. Electron loss to the surface is related to electron temperature because it is a function of ambipolar diffusion, which scales with $1+T_{\rm e}/T$.

Most past simulations of hypersonic plasma flows assumed that the free-electron-vibrational relaxation is an infinitely fast process and, thus, that a single transport equation could be used to obtain both the vibrational temperature and the electron temperature as in \cite{nasa:1989:gnoffo} for instance. There are relatively few previous numerical simulations of high-speed flows that solved the electron temperature in non-equilibrium (see, for instance, \cite{jtht:2012:kim, jtht:2013:farbar, jtht:1991:candler, jsr:2003:josyula, aiaaj:2022:sawicki}). In all such studies, the elastic and inelastic electron energy relaxation terms were taken from \cite{aiaaconf:1984:lee, aiaaconf:1985:lee}, who based his theoretical expressions on the experimentally determined cross-sections obtained by  \cite{nrl:1982:slinker}. There are several issues with Lee's approach to modeling the electron energy transport relaxation terms. The first is that it requires accurate cross-sections for all inelastic processes. Thus, one set of cross-sections is needed for the electron impact excitation of the first vibrational energy level, another set for the second vibrational energy level, another set for the electron impact excitation of the rotational energy, another set for Townsend ionization, and so on. But even if all the cross-sections of all the electron impact processes could be obtained with high accuracy, the electron energy relaxation source terms could still be tainted with significant error. This is because the cross-sections need to be converted to reaction rates function of electron temperature. This conversion is nontrivial and frequently introduces considerable errors.

This paper proposes a new form of the electron energy transport source terms that overcomes the deficiencies of Lee's approach. The source terms are first rewritten so that they correspond to the sum of energy relaxation terms due to elastic collisions with charged and neutral species and energy relaxation terms due to inelastic collisions with the neutral species. Analytical expressions are then derived for the electron-ion and electron-neutral elastic collisions. We then show that all electron energy losses due to inelastic collisions with neutral species can be written in one term function of only two species-dependent parameters: the reduced electric field and the reduced electron mobility. The resulting electron energy relaxation source terms presented herein are advantageous over previous formulations by being simpler to implement and more accurate when experimental data of the reduced electric field and reduced mobility are available (as is the case for most neutral air species). It is emphasized that the proposed approach accounts for all inelastic electron energy relaxation processes without requiring the error-prone cross-sections and/or rates of each inelastic process. 

Previous work at expressing the electron energy losses as a function of the reduced electric field by \cite{misc:1995:boeuf} assume that the plasma has only three components (electrons, ions, and one type of neutral molecule) and further assume that the electron energy gains from collisions (with either neutrals or ions) are negligible. This formulation is inadequate for hypersonic plasma flows because the latter typically involve more than one neutral species and often lead to an electron temperature lower than the gas temperature, thus rendering electron energy gains significant. In contrast, the proposed approach is suitable for hypersonic plasmas with various types of neutral molecules, where the electron temperature can be either lower or higher than the gas temperature.

To assess the validity of the proposed electron energy transport equation, we simulate cases for which a comparison with experimental data is available, such as the RAM-C-II flowfield, a glow discharge in a hypersonic boundary layer, and a uniformly applied electric field on a plasma flow. The glow discharge test case includes a non-neutral sheath of significant size that can impact the electron temperature. Additionally, the RAM-C-II flight test also includes non-neutral Debye sheaths that can affect electron temperature through electron cooling as first observed by \cite{pf:2021:parent}. Therefore, it is important that the mass, momentum, and energy transport equations include proper modeling of the plasma sheaths and their effects on electron temperature.

\section{Mass, Momentum, and Energy Transport Equations}

To accurately capture either Debye sheaths or cathode sheaths, the velocity of the charged species is obtained from a momentum equation that includes body forces due to the electric field and collision forces with neutral and charged species. Neglecting inertia effects but retaining the pressure gradient effects in finding the velocity difference between the species and the bulk leads to the so-called 'drift-diffusion model' outlined in \cite{book:2022:parent}.  

The $i$th component of the velocity of the $k$th species, $V_i^k$, can be expressed as the sum of the bulk (mixture) flow velocity $\vec{V}$, drift terms (which are function of the electric field $\vec{E}$), and diffusion terms (which are function of either the partial pressure gradient $\partial P_k/\partial x$ or the mass fraction gradient $\partial w_k/\partial x$):
\begin{equation}
  V^{k}_i = \left\{
  \begin{array}{ll}\mfd
  V_i+s_k \mu_k  {E}_i
             -  \frac{\mu_k}{|C_k| N_k} \frac{\partial P_k}{\partial x_i} & \textrm{for electrons and ions} \alb\mfd
  V_i - \frac{\nu_k}{\rho_k} \frac{\partial w_k}{\partial x_i} & \textrm{for neutrals}
  \end{array}
  \right.
\label{eqn:Vk}
\end{equation}
In the latter, $C_k$ represents the charge of the $k$th species, $s_k$ denotes the species charge sign (-1 for negatively charged species or +1 otherwise), $\nu_k$ is the mass diffusion coefficient, and $\mu_k$ represents the mobility of the $k$th species. Additionally, the partial pressure $P_k$ is obtained from Dalton's law of partial pressure as follows:
\begin{equation}
P_k= \left\{ 
\begin{array}{ll}
N_{\rm e} k_{\rm B}  T_{\rm e} & \textrm{for electrons} \alb
N_{k} k_{\rm B}  T & \textrm{for ions and neutrals} 
\end{array}
\right.
\label{eqn:Pk}
\end{equation}
with $T$ the gas temperature, $T_{\rm e}$ the electron temperature, $N_k$ the number density of the $k$th species, and $k_{\rm B}$ the Boltzmann constant. 

Given the species velocity, we can express the mass conservation equation for the $k$th species (whether neutral or charged) as follows:
\begin{equation}
\frac{\partial}{\partial t} \rho_k + \sum_i \frac{\partial }{\partial x_i}\rho_{k} V_i^{k} = W_{k}  
\label{eqn:masstransport}
\end{equation}
where $\rho_k$ is the mass density of the $k$th species and where $W_k$ is the chemical source term consisting of the creation and destruction of the $k$th species due to chemical reactions.

The momentum equation for the bulk of the flow includes the Navier-Stokes viscous stresses $\tau_{ji}$ as well as source terms to account for the electric field force on the non-neutral plasma:
\begin{equation}
  \rho \frac{\partial V_i }{\partial t}+ \sum_{j=1}^3 \rho V_j \frac{\partial V_i}{\partial x_j}
=
-\frac{\partial P}{\partial x_i} 
+ \sum_{j=1}^3 \frac{\partial \tau_{ji}}{\partial x_j}
+ \rho_{\rm c}{E}_i
\label{eqn:momentum}
\end{equation}
where $\rho_{\rm c}=\sum_k C_k N_k$ is the net charge density and where $P=\sum_k P_k$ is the sum of the partial pressures including the electron partial pressure.

\begin{figure}[t]
     \centering
     \includegraphics[width=0.35\textwidth]{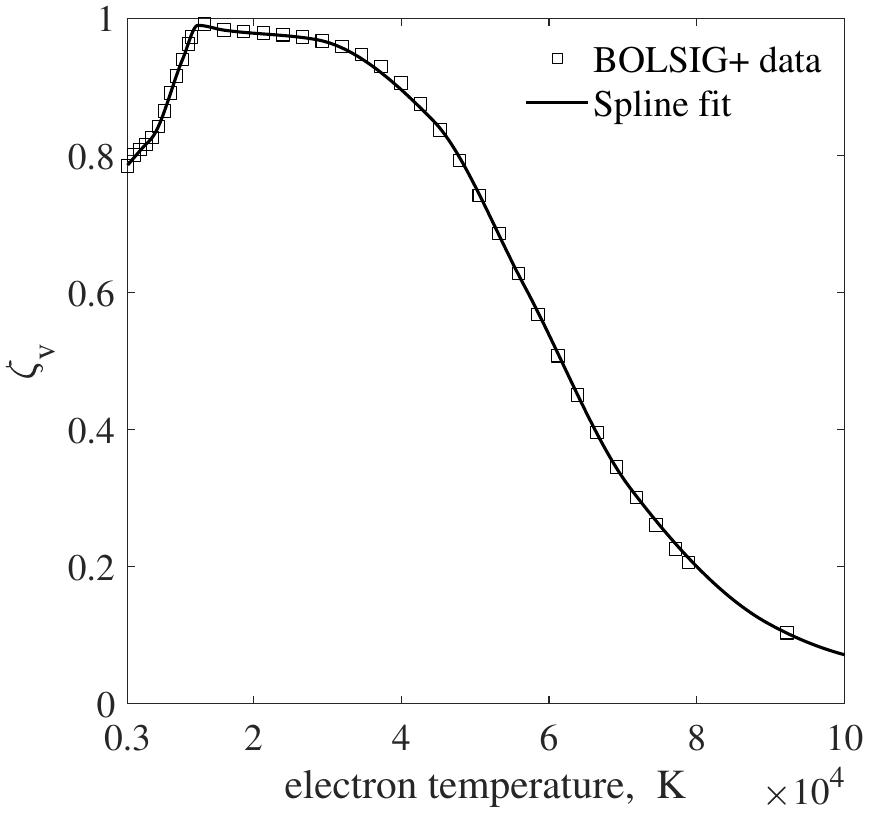}
     \figurecaption{     
Fraction of electron Joule heating consumed in the excitation of the  $\rm N_2$ vibrational energy.}
     \label{fig:zetav_Te_spline}
\end{figure}

 Energy transport equations are needed to determine translational temperature, nitrogen vibrational temperature and electron temperature in non-equilibrium. We here limit non-equilibrium effects of the vibrational energy to the molecular nitrogen species because, for the hypersonic test cases here considered, the majority of the molecules are N$_2$ and  only a few percent are NO or O$_2$. For molecular species other than N$_2$, the vibrational temperature is assumed to be equal to the translational and rotational temperature. 
 The energy transport equation for the nitrogen vibrational temperature $T_{\rm v}$ is taken from \cite{aiaa:2001:macheret,aiaaconf:1999:macheret}:
\begin{equation}
 \begin{array}{l}
  \mfd\frac{\partial}{\partial t} \rho_{\rm N_2} e_{\rm v}
     + \sum_{j=1}^{3} \frac{\partial }{\partial x_j}
       \rho_{\rm N_2} V_j e_{\rm v}
     - \sum_{j=1}^{3} \frac{\partial }{\partial x_j} \left(
            \kappa_{\rm v}  \frac{\partial T_{\rm v}}{\partial x_j}\right)\alb\mfd
     - \sum_{j=1}^{3} \frac{\partial }{\partial x_j} \left(
            e_{\rm v} \nu_{\rm N_2}  \frac{\partial w_{\rm N_2}}{\partial x_j}\right)
 = 
  \frac{N_{\rm N_2}}{N} \zeta_{\rm v} Q_{\rm J}^{\rm e}   + {\frac{\rho_{\rm N_2}}{\tau_{\rm vt}}}\left( e_{\rm v}^0 -e_{\rm v} \right) + W_{\rm N_2} e_{\rm v}
\end{array}
\label{eqn:vibrationalenergy}
\end{equation}
where $e_{\rm v}$ is the nitrogen vibrational energy, $e_{\rm v}^0$ is the nitrogen vibrational energy at equilibrium, $\kappa_{\rm v}$ is the nitrogen vibrational thermal conductivity, $N$ is the total number density of the mixture, and $\tau_{\rm vt}$ is the vibration-relaxation time taken from \cite{jpp:2007:parent} and \cite{aiaa:2001:macheret}. Following the derivation outlined in the Appendix of \cite{aiaa:2016:parent}, the electron Joule heating term $Q_{\rm J}^{\rm e}$ can be shown to be equal to:
\begin{equation}
Q_{\rm J}^{\rm e}=\frac{|C_{\rm e}| N_{\rm e}}{\mu_{\rm e}} \left|\vec{V}_{\rm e}-\vec{V} \right|^2
\end{equation}
As well, $\zeta_{\rm v}$ represents the fraction of electron Joule heating consumed by exciting the nitrogen vibrational energy levels. To obtain $\zeta_{\rm v}$, first note that the electron Joule heating in a uniform plasma is proportional to  $(E^\star)^2$ with the reduced electric field $E^\star\equiv |\vec{E}|/N$. Therefore the fraction of Joule heating that goes directly to the vibrational energy levels can be written as $\zeta_{\rm v}=(E^\star_{\rm vib})^2/(E^\star)^2 = 1-({E^\star_{\rm novib}}/E^\star)^2$. Here, $E^\star_{\rm novib}$ is obtained with BOLSIG+ using cross sections for all collision processes but excluding the vibrational excitation process, while $E^\star$ is obtained by including all such processes, i.e. elastic, rotational, vibrational, electronic excitation, and ionization. The  cross sections are sourced from the database compiled by \cite{pcpp:1992:morgan}. A spline curve-fit based on the latter BOLSIG+ calculations  is shown in Fig.\ \ref{fig:zetav_Te_spline}, with the data points needed to construct the \cite{sjna:1980:fritsch} monotone cubic spline  listed in Table\ \ref{tab:zetav_table_spline}.

\begin{table*}[!ht]
\fontsizetable
\begin{center}
\begin{threeparttable}
\tablecaption{Spline control points giving the fraction of the electron Joule heating consumed in the excitation of the N$_2$ vibrational energy.}
    \begin{tabular*}{0.99\textwidth}{@{}l@{\extracolsep{\fill}}l@{}}
    \toprule
    $T_{\rm e}$, K & $\zeta_{\rm v}$  \\
        \midrule
   \begin{minipage}[t]{0.52\textwidth}\raggedright  
3000,	5053,	6437,	10451,	12486,	16037,	28431,	44364,	56769,	70927,	89529,	100159,	119643,	150627,	186021,	232091,	345815,	630010,	3000000
 \end{minipage}  & \begin{minipage}[t]{0.48\textwidth}\raggedright 
0.7859,	0.8115,	0.828,	0.9411,	0.9895,	0.9825,	0.9689,	0.8501,	0.6073,	0.3154,	0.1178,	0.0709,	0.0321,	0.012,	0.0039,	0.0015,	0.001,	0.0006,	0.0
\end{minipage}\alb
    \bottomrule
    \end{tabular*}
\label{tab:zetav_table_spline}
\end{threeparttable}
\end{center}
\end{table*}

To determine the electron temperature in non-equilibrium, the following energy transport equation is derived from the first law of thermodynamics:
\begin{equation}
 \frac{\partial }{\partial t}\rho_{\rm e} e_{\rm e} + \sum_{j=1}^3  \frac{\partial }{\partial x_j} \left( \rho_{\rm e} V^{\rm e}_j h_{\rm e}  - \kappa_{\rm e} \frac{\partial T_{\rm e}}{\partial x_j}  
\right)
= 
 W_{\rm e} e_{\rm e}
+   C_{\rm e} N_{\rm e} \vec{E} \cdot \vec{V}_{\rm e}  
- Q_{\rm e}
\label{eqn:electronenergy}
\end{equation}
with $e_{\rm e} =\frac{3}{2} \frac{k_{\rm B}}{m_{\rm e}} T_{\rm e}$ the electron specific energy, $h_{\rm e}=\frac{5}{2} \frac{k_{\rm B}}{m_{\rm e}} T_{\rm e}$ the electron specific enthalpy, and $\kappa_{\rm e}=\frac{5}{2} N_{\rm e} k_{\rm B}^2 T_{\rm e} \mu_{\rm e} /|C_{\rm e}|$ the electron thermal conductivity (see \cite{ht:1997:raizer}). As well, $Q_{\rm e}$ is the electron energy losses due to elastic or inelastic collisions with the heavy particles and which will be described in detail in the next section. No kinetic energy terms appear in the electron energy equation because the inertia terms are neglected within the charged species momentum equations.

The total energy transport equation is derived by summing all the energy equations (including electron energy, N$_2$ vibrational energy, ions total energy, and neutrals total energy). This yields the following:
\begin{equation}
\begin{array}{l}\mfd
 \frac{\partial }{\partial t}\rho e_{\rm t}
+ \sum_{j=1}^3  \frac{\partial }{\partial x_j} V_j \left(\rho  e_{\rm t} +  P \right)
 \alb\mfd
- \sum_{j=1}^3  \frac{\partial }{\partial x_j} \left(
   \nu_{\rm N_2} e_{\rm v}\frac{\partial w_{\rm N_2}}{\partial x_j} 
  + \sum_{k=1}^\ns  \rho_k (V^k_j-V_j) {(h_k+h_k^\circ)}  
\right)
 \alb\mfd
-\sum_{i=1}^{3}\frac{\partial }{\partial x_i}\left((\kappa_{\rm n}+\kappa_{\rm i}) \frac{\partial T}{\partial x_i} 
+ \kappa_{\rm v} \frac{\partial T_{\rm v}}{\partial x_i} +\kappa_{\rm e} \frac{\partial T_{\rm e}}{\partial x_i}\right)
 \alb\mfd
=
 \sum_{i=1}^3 \sum_{j=1}^3  \frac{\partial }{\partial x_j} \tau_{ji} V_i
+ \vec{E}\cdot\vec{J}
\end{array}
\label{eqn:totalenergy}
\end{equation}
with the current density vector $\vec{J} \equiv \sum_{k}  C_k N_k  \vec{V}_k$ and with $P$ the sum of the partial pressures.   The total specific energy $e_{\rm t}$ is defined as the sum of the kinetic energy of the flow, the nitrogen vibrational energy, the internal energy of each species $e_k$, as well as the heat of formation of each species $h_k^\circ$: 
\begin{equation}
e_{\rm t} \equiv w_{\rm N_2} e_{\rm v} + \sum_{k=1}^\ns w_k (e_k + h_k^\circ) +\frac{1}{2} \vec{V}^2
\label{eqn:specifictotalenergy}
\end{equation}
where $e_k$ is the specific internal energy obtained from the specific enthalpy as $e_k=h_k-P_k/\rho_k$.
The enthalpy of the $k$th species excluding the heat of formation is denoted as $h_k$. Note that $h_{\rm N_2}$ does not include the vibrational energy. The \cite{nasa:2002:mcbride} high temperature polynomials  are used to determine the enthalpies (including the heat of formation) while either the  \cite{nasa:1990:gupta} model or the  \cite{book:1984:dixon-lewis} model is used to find the viscosity $\eta$, the thermal conductivity of the neutrals $\kappa_{\rm n}$, the thermal conductivity of the ions $\kappa_{\rm i}$, the mass diffusion coefficients $\nu_k$, and the mobilities $\mu_k$. Although they do not explicitly appear in the equations, the electronic and non-N$_2$ vibrational energies are included in the energy balance because they are part of the \cite{nasa:2002:mcbride} enthalpies. In doing so, it is assumed that the electronic temperature and the non-N$_2$ vibrational temperature is equal to the translational and rotational temperature.

To close the latter transport equations, we need an equation for the electric field which can be obtained from Gauss's law as follows:
\begin{equation} 
\sum_{i=1}^3 \frac{\partial }{\partial x_i}\left(\epsilon_r \frac{\partial \phi}{\partial x_j}\right)=-\frac{\rho_{\rm c}}{\epsilon_0}   \label{eqn:gauss_potential} 
\end{equation}
with $\phi$, $\epsilon_0$, and $\epsilon_{\rm r}$   the electric field potential, the permittivity of free space, and the relative permittivity, respectively. From the potential we can find the electric field by taking the negative of its gradient (i.e.\ $\vec{E}=-\vec{\nabla} \phi$).

Regarding the surface boundary conditions, we assume that there is no surface catalysis for any neutral species. For charged species, however, secondary electron emission and surface catalysis occur when electrons and ions recombine at the surface. These effects can be represented through the following boundary conditions for the electron and ion densities at solid surfaces:
\begin{equation}
\frac{\partial }{\partial \chi} N_+ V^{+}_\chi = 0
{~~\rm and~~}
N_{-}=0
{~~\rm and~~}
N_{\rm e}=\frac{\gamma_{\rm e}}{\mu_{\rm e}} \sum_{k=1}^{n_{\rm s}} N_k \mu_k \beta_k^+
{~~\rm for~}
E_\chi<0
\end{equation}
\begin{equation}
N_{+}=0
{~~~~~\rm and~~~~~}
\frac{\partial }{\partial \chi} N_- V^{-}_\chi = 0
{~~~~~\rm and~~~~~}
\frac{\partial }{\partial \chi} N_{\rm e} V^{\rm e}_\chi= 0
{~~~~~\rm otherwise} 
\end{equation}
In the latter, we use  ``+'', ``-'', and ``e'' to represent positive ions, negative ions, and electrons, respectively. Additionally, $\beta_k^+$ is equal to 1 when the $k$th species is a positive ion and to zero otherwise, and $\chi$ denotes a coordinate perpendicular to the surface, oriented away from the surface and toward the fluid. It is noted that when the electric field points towards the surface ($E_\chi<0$), the latter surface boundary condition for the electrons assumes for every ion impinging the surface there are $\gamma_{\rm e}$ electrons released from the surface. As recommended by \cite{pre:2000:hagelaar}, the boundary condition for the electrons further assumes that all electrons emitted from the surface do so through drift only and that all ions impinging the surface also do so through drift only. This is, generally, an excellent approximation for cathodes and a fair approximation for dielectrics. Further, on all dielectric or conductor surfaces, the derivative of the electron temperature normal to the surface is fixed to zero. 

The secondary electron  emission coefficient, $\gamma_{\rm e}$, is assigned a value of 0.1 for dielectrics and 0.6--0.8 for electrodes.  Although this is slightly higher than the expected range of 0.1--0.2 often used for electrodes, we note that  $\gamma_{\rm e}$ is not well known on metal surfaces except when the incoming ion energy is more than 1~keV (as in \cite{prb:1979:baragiola} for instance). This is an order of magnitude higher than the ion energies directed toward the cathode observed in the glow discharge test case presented below. Additionally, \cite{psst:1999:phelps} observed that incoming neutrals could release electrons from the surface. Incoming energetic electrons were also observed by \cite{misc:1956:kollath} and \cite{misc:1958:dekker} to release electrons from the surface. Because we do not have a separate process for secondary electron emission from neutral-electron or electron-electron collisions,  $\gamma_{\rm e}$ needs to account for all processes that result in secondary electron emission. This includes ion-electron, electron-electron, and neutral-electron collisions at the surface. Consequently, our \emph{effective} electron yield per ion, $\gamma_{\rm e}$, is higher than it would be if it would only account for ion-electron surface collisions.

\section{Proposed electron energy relaxation model}

Within the electron energy transport equation shown in Eq.\ (\ref{eqn:electronenergy}), all the electron energy collisional losses (with either ions or neutrals and including both elastic and inelastic collisions) were incorporated within a single term denoted as $Q_{\rm e}$. This term is referred to here as the electron energy relaxation term. We aim to find an expression for $Q_{\rm e}$ that is well-suited for simulating discharges as well as high-temperature hypersonic flows in thermal non-equilibrium.

We first rewrite $Q_{\rm e}$ as a sum of electron energy losses-gains due to elastic collisions and electron energy losses due to inelastic collisions as follows:
\begin{equation}
Q_{\rm e}=Q_{\rm elastic}+Q_{\rm inelastic}
\end{equation}
where $Q_{\rm elastic}$ are the electron energy losses-gains due to elastic collisions and $Q_{\rm inelastic}$ are the electron energy losses due to inelastic collisions.

\subsection{Electron energy source terms due to elastic collisions with ions and neutrals}

Assuming that few collisions result in electron loss through attachment or recombination, the electron energy gains/losses with all heavy particles (either neutrals or ions) due to elastic collisions can be obtained following \cite[page 16]{book:1991:raizer} and
 \cite{jfm:1964:appleton}:
\begin{equation}
Q_{\rm elastic}
= 
  \sum_k^{k\ne {\rm e}}  3 N_{\rm e} k_{\rm B} (T_{\rm e}-T) \frac{ m_{\rm e}}{m_k}   \nu_{{\rm e}k}  
\end{equation}
where $\nu_{{\rm e}k}$ is the collision frequency between the electrons and species $k$ and $m_k$ is the mass of one particle of the $k$th species.  We can rewrite the collision frequency as function of the mobility as in \cite[page 9]{book:1991:raizer} or the collision cross section in the following form:
\begin{equation}
\nu_{{\rm e}k} = \frac{|C_{\rm e}|}{m_{\rm e} \mu_{{\rm e}k}} = N_k \sigma_{{\rm e}k} \overline{q_{\rm e}} 
\label{eqn:collisionfrequency}
\end{equation}
where $\mu_{{\rm e}k}$ is the electron mobility within species $k$ and where  $\sigma_{{\rm e}k}$ is the collision cross section between an electron and species $k$ and $\overline{q_{\rm e}}$ is the  electron thermal velocity. It is convenient (for reasons that will be outlined later) to express the collision frequency as a function of mobility for the electron-neutral collisions and as a function of the collision cross section for the electron-ion collisions. This leads to $Q_{\rm elastic}$ being split into two terms, namely:
\begin{equation}
Q_{\rm elastic}
= 
  \sum_k  \frac{3 \beta_k^{\rm n} k_{\rm B} |C_{\rm e}| \rho_{\rm e} \rho_k  (T_{\rm e}-T)}{m_{\rm e} m_k^2 (\mu_{\rm e}N)_k}        
 + \sum_k \beta_k^{\rm i} 3 \rho_{\rm e} k_{\rm B} (T_{\rm e}-T) \overline{q_{\rm e}} \frac{N_k}{m_k} \sigma_{ek} 
\end{equation}
where $\beta_k^{\rm i}$ is equal to 1 when the $k$th species is an ion and to 0 otherwise. Similarly, $\beta_k^{\rm n}$ is equal to 1 when the $k$th species is a neutral and to 0 otherwise.
We can further express  the collision cross-section following \cite[page 58]{book:2004:lieberman}:
\begin{equation}
\sigma_{{\rm e}k}= \frac{8}{\pi} b_0^2 \ln \Lambda
\end{equation}
with $\ln \Lambda$ the Coulomb logarithm which can be found in \cite[p. 14]{book:1991:raizer}, or in \cite{nasa:1990:gupta}, or in the Navy Research Laboratory (NRL) Formulary by \cite[page 34]{nrl:2002:huba} for instance. We prefer the Coulomb logarithm listed in the NRL formulary because, when used for the electron-ion collisions needed to determine electron mobility, this leads to an electrical conductivity closer to the experimental data outlined in \cite{tvt:1969:asinovskiy} and in \cite{book:spitzer:1956} when the air plasma is strongly ionized and its temperature is more than 7,000~K. Thus, the Coulomb logarithm we recommend is the one from \cite[page 34]{nrl:2002:huba}:
\begin{equation}
\ln \Lambda=23-\ln \left(N_{\rm e}^{0.5} T_{\rm e}^{-1.5}\right)
\end{equation}
with $T_{\rm e}$ in eV and $N_{\rm e}$ in cm$^{-3}$.
 As well, following \cite[page 56]{book:2004:lieberman}, the classical distance of closest approach $b_0$ is equal to:
\begin{equation}
b_0 = \frac{|C_{\rm i}| |C_{\rm e}|}{4 \pi \epsilon_0 } \frac{1}{\frac{1}{2} m_{\rm e} \overline{q_{\rm e}}^2}
\end{equation}
with $\overline{q_{\rm e}}$ the average speed of the electrons approaching an ion which we take equal to the electron thermal velocity because the latter is typically orders of magnitude greater than the ion thermal velocity (unless the electron temperature is less than $10^{-4}T$ or so, which is unlikely to be the case):
\begin{equation}
\overline{q_{\rm e}} = \sqrt{\frac{8 k_{\rm B} T_{\rm e}}{\pi m_{\rm e}}}
\label{eqn:electronthermalvelocity}
\end{equation}
Then the electron energy loss-gain due to elastic collisions with both neutrals and ions becomes:
\begin{align}
Q_{\rm elastic}
&= 
  \sum_k  \frac{3 \beta_k^{\rm n} k_{\rm B} |C_{\rm e}| N_{\rm e} N_k  (T_{\rm e}-T)}{ m_k (\mu_{\rm e}N)_k}\nonumber\\        
 &+ \sum_k \beta_k^{\rm i}  N_{\rm e} N_k  (T_{\rm e}-T )  \frac{6 k_{\rm B} C_{\rm i}^2 C_{\rm e}^2 \ln \Lambda}{ \pi^3 \epsilon_0^2 m_{\rm e} m_k \overline{q_{\rm e}}^3}  
\label{eqn:Qelastic}
\end{align}
In the latter, $(\mu_{\rm e} N)_k$ corresponds to the reduced electron mobility in species $k$ and can be obtained for each neutral species from experiments or from cross-sectional data and a Boltzmann equation solver.

\subsection{Electron energy source terms due to inelastic collisions with neutrals}

Here, we derive a relationship between the inelastic electron energy loss terms with the neutrals and the reduced electric field. To achieve this, let us consider the energy transport equation for a steady-state plasma with uniform properties (no spatial gradients):
\begin{equation}
0
= 
 W_{\rm e} e_{\rm e}
+   C_{\rm e} N_{\rm e} \vec{E} \cdot \vec{V}_{\rm e}  
- Q_{\rm e} 
\end{equation}
However, from the electron mass conservation equation (\ref{eqn:masstransport}), it follows that $W_{\rm e}=0$ at steady-state when there are no spatial gradients of the electron density:
\begin{equation}
0
= 
   C_{\rm e} N_{\rm e} \vec{E} \cdot \vec{V}_{\rm e}  
- Q_{\rm e} 
\end{equation}
Imposing a steady-state with $W_{\rm e}=0$ does not imply that there is no electron destruction or production through chemical reactions. It rather implies that the rate of electron production equals the rate of electron destruction, as is necessarily the case in a steady-state uniform plasma. We proceed in this manner because the experimental data (or BOLSIG+ data) of electron energy losses that will be used to close our model are obtained for a uniform and steady plasma.    Nonetheless, our so-derived model can be deployed to non-uniform and/or unsteady  plasmas as long as the electron energy relaxation time is less than the time scales of the fastest macroscopic processes predicted by the fluid transport equations. Such will be discussed in more detail for the glow discharge test case presented below.

Isolate $Q_{\rm e}$ and note that $C_{\rm e}$ is the negative of the elementary charge and that the electron velocity in a gas at rest and with no spatial gradients corresponds to $\vec{V}_{\rm e}= - \mu_{\rm e} \vec{E}$:
\begin{equation}
Q_{\rm e}
= |C_{\rm e}| N_{\rm e} \mu_{\rm e} \vec{E}^2   
\end{equation}
Also, we can define the reduced electric field  as $E^\star\equiv |\vec{E}|/N$ and the reduced mobility as $\mu_{\rm e}^\star\equiv \mu_{\rm e} N$. Then the latter becomes:
\begin{equation}
Q_{\rm e}
=  |C_{\rm e}| N_{\rm e} N \mu_{\rm e}^\star (E^\star)^2   
\end{equation}
Recall that $Q_{\rm e}$ corresponds, by definition, to the sum of the electron energy loss due to elastic and inelastic collisions. Furthermore, we can add and subtract the term $\left(E^\star_{\rm elastic}\right)^2$ from the reduced electric field. Then, we obtain:
\begin{equation}
\begin{array}{l}
Q_{\rm inelastic} + Q_{\rm elastic} \alb\mfd
~~=  |C_{\rm e}| N_{\rm e} N \mu_{\rm e}^\star \left( (E^\star)^2 - \left(E^\star_{\rm elastic}\right)^2\right) 
+ |C_{\rm e}| N_{\rm e} N \mu_{\rm e}^\star \left(E^\star_{\rm elastic}\right)^2    
\end{array}
\end{equation}
Should we define the reduced electric field responsible for elastic collisions as:
\begin{equation}
 \left(E^\star_{\rm elastic}\right)^2 \equiv \frac{Q_{\rm elastic}}{|C_{\rm e}| N_{\rm e} N \mu_{\rm e}^\star}
 \label{eqn:Estarelasticdefinition}
\end{equation}
and after isolating the electron energy loss due to inelastic collisions, we obtain:
\begin{equation}
Q_{\rm inelastic}  
=  |C_{\rm e}| N_{\rm e} N \mu_{\rm e}^\star \left( (E^\star)^2 - \left(E^\star_{\rm elastic}\right)^2\right) 
\label{eqn:Qinelastic1}
\end{equation}
The terms on the right-hand-side can be easily found for the case of a weakly-ionized 3-component plasma with one type of neutral molecules, one type of ions, and electrons. Indeed, we can find $\mu_{\rm e}^\star$ and $E^\star$ as a function of $T_{\rm e}$ from experiments or from given cross sections and a Boltzmann equation solver.  In such a scenario, the electron energy losses due to Coulomb collisions are negligible and $E^\star_{\rm elastic}$ becomes function of only the electron energy losses due to electron-neutral elastic collisions. The only remaining term on the right-hand-side that needs to be determined is $E^\star_{\rm elastic}$. This can be found by substituting the elastic collisions between electrons and neutrals in Eq.~(\ref{eqn:Qelastic}) into Eq.~(\ref{eqn:Estarelasticdefinition}):
\begin{equation}
 \left(E^\star_{\rm elastic}\right)^2 =
   \frac{3  k_{\rm B}    (T_{\rm e}-T_{\rm ref})}{ m_{\rm n} (\mu_{\rm e}^\star)^2}      
 \label{eqn:Estarelasticneutral}
\end{equation}
with $m_{\rm n}$ the mass of a neutral particle and $T_{\rm ref}$ the reference gas temperature in the experiments used to obtain the $E^\star(T_{\rm e})$ data. Because $\mu_{\rm e}^\star$ is a sole function of electron temperature in this case, it follows that $E^\star_{\rm elastic}$ for electron-neutral collisions is function solely of constants (Boltzmann constant and particle mass), of the reference gas temperature, and of the electron temperature.   

The expression for the inelastic electron energy losses, as shown in Eq.~(\ref{eqn:Qinelastic1}), can be generalized for a gas mixture involving multiple neutral species by noting that the energy loss processes of each neutral species scale with the corresponding molar fraction. After substituting the elastic electric field from Eq.~(\ref{eqn:Estarelasticneutral}), this yields: 
\begin{equation}
Q_{\rm inelastic}  
=  \sum_k \beta_k^{\rm n} |C_{\rm e}| N_{\rm e} N_k (\mu_{\rm e}^\star)_k \left( (E^\star_k)^2 -  \frac{3  k_{\rm B}    (T_{\rm e}-T_{\rm ref})}{ m_{k} (\mu_{\rm e}^\star)^2_k} \right) 
\label{eqn:Qinelastic}
\end{equation}
where $\beta_k^{\rm n}$ is 1 when the $k$th species is a neutral and to 0 otherwise. 

It is instructive to compare Eq.~(\ref{eqn:Qinelastic}) to the inelastic electron energy losses due to electron impact collisions with the neutrals written in the standard form:
\begin{equation}
Q_{\rm inelastic}  
=   \sum_k   \beta_k^{\rm n} N_{\rm e} N_k  \sum_l k_{kl} \mathcal{E}_{kl}
\label{eqn:Qinelasticstandard}
\end{equation}
where $l$ denotes an electron impact process, $k_{kl}$ is the rate coefficient of the $l$th electron impact process acting on the $k$th neutral species, and $\mathcal{E}_{kl}$ is the activation energy of the $l$th electron impact process of the $k$th species. Equating Eqs.~(\ref{eqn:Qinelastic}) and (\ref{eqn:Qinelasticstandard}) we obtain the following  expression for the sum of the inelastic rate coefficients multiplied by the activation energy for the $k$th neutral species:
\begin{equation}
  \sum_l k_{kl} \mathcal{E}_{kl}  
=  |C_{\rm e}|  \left(\mu_{\rm e}^\star\right)_k \left( (E^\star_k)^2 - \frac{3  k_{\rm B}    (T_{\rm e}-T_{\rm ref})}{ m_{k} (\mu_{\rm e}^\star)^2_k}\right) 
\label{eqn:suminelasticenergies}
\end{equation}

The right-hand-side depends only on electron temperature, a reference gas temperature (a fixed constant), and only two species-dependent parameters: the reduced electric field and the reduced mobility. Because  both the reduced mobility and the reduced electric field are only function of electron temperature, it follows that the proposed model yields a sum of the electron impact rates times the activation energy which depends only on electron temperature. 

\begin{figure}[t]
     \centering
     \subfigure[]{\includegraphics[width=0.35\textwidth]{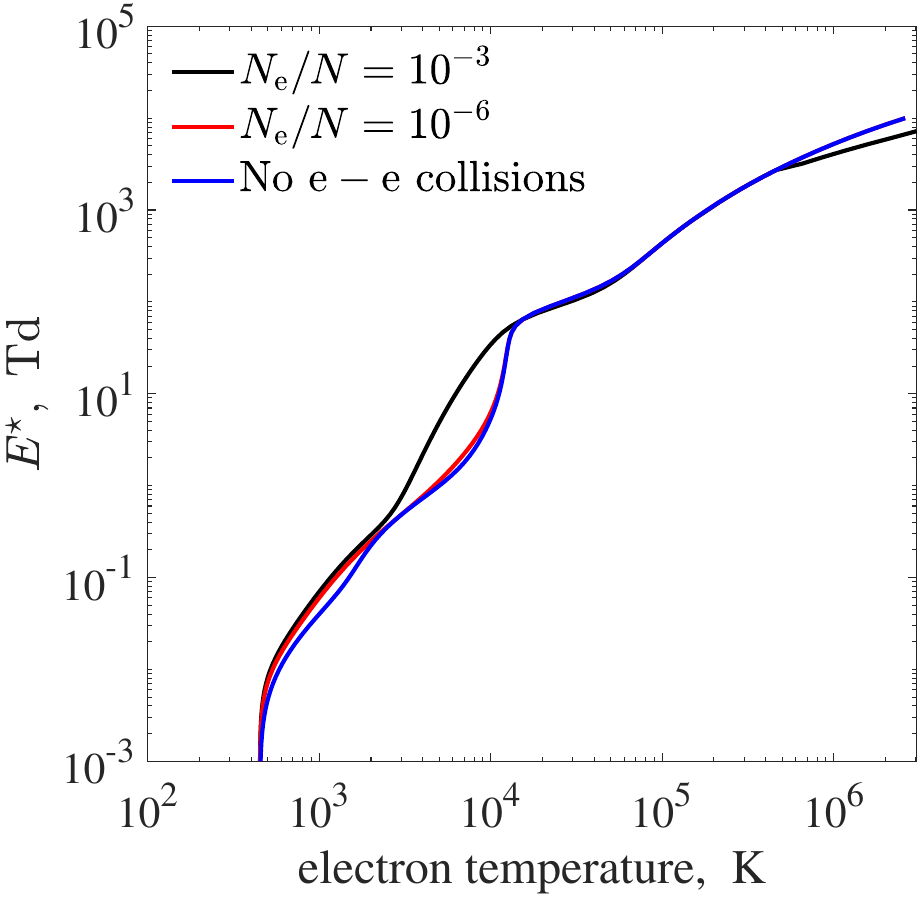}}
     \subfigure[]{\includegraphics[width=0.35\textwidth]{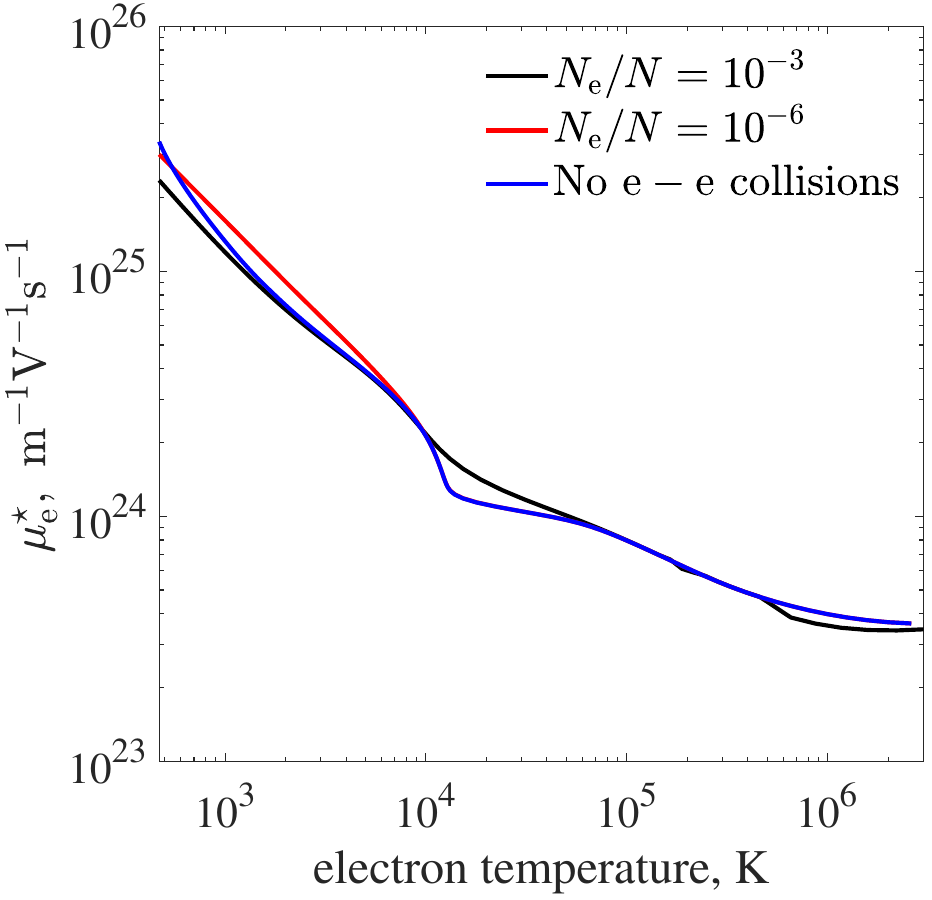}}
     \figurecaption{Effect of plasma density on (a) the reduced electric field and (b) the reduced electron mobility in $\rm N_2$ with number density $10^{24}~\rm m^{-3}$.}
     \label{fig:ee_collisions_N2}
\end{figure}

It is important to note that while the reduced electric field and reduced mobility are typically determined from experiments taking place in a weakly-ionized plasma, in which electron-ion and electron-electron collisional losses are negligible, they generally remain accurate when deployed to strongly-ionized plasmas. For instance, in Fig.~\ref{fig:ee_collisions_N2}, we show the impact of a change in plasma density on the reduced electric field and the reduced mobility. The results are obtained using BOLSIG+ including electron-electron collisions and the Morgan cross sections found in the LXCat database by \cite{ppp:2017:pitchford}. Clearly,  a change in  the ionization fraction from $10^{-6}$ to $10^{-3}$ leads to small differences in the electron energy inelastic losses except when the electron temperature is in the range 3,000--10,000~K. This is generally not a concern, as plasmas with an electron molar fraction greater than 0.001 are likely to either (i) have an electron temperature above 10,000 K or (ii) maintain equilibrium between the electron and gas temperatures. However, if an N$_2$ plasma is highly ionized, in thermal non-equilibrium, and has an electron temperature within 0.3--1~eV, it is observed that the reduced electric field and mobility (and consequently, the inelastic electron energy losses) are significantly influenced by electron-electron collisions. Therefore, in that electron temperature range, they depend not only on the electron temperature but also on the plasma density.

\subsection{Electron energy source terms including both elastic and inelastic collisions}

Let us now add the electron energy losses due to elastic collisions with ions and neutrals shown in Eq.~(\ref{eqn:Qelastic}) to the electron energy losses due to inelastic collisions with the neutrals shown in Eq.~(\ref{eqn:Qinelasticstandard}). 
\begin{equation}
\begin{array}{l}\mfd
Q_{\rm e}
= \sum_k  \beta_k^{\rm n} N_{\rm e} N_k  \sum_l k_{kl} \mathcal{E}_{kl}
+  \sum_k  \frac{3 \beta_k^{\rm n} k_{\rm B} |C_{\rm e}| N_{\rm e} N_k  (T_{\rm e}-T)}{ m_k (\mu_{\rm e}^\star)_k}\alb\mfd~~~ 
 + \sum_k \beta_k^{\rm i}  N_{\rm e} N_k  (T_{\rm e}-T )  \frac{6 k_{\rm B} C_{\rm i}^2 C_{\rm e}^2 \ln \Lambda}{ \pi^3 \epsilon_0^2 m_{\rm e} m_k \overline{q_{\rm e}}^3}  
 \end{array}
\end{equation}
Let us then use Eq.~(\ref{eqn:suminelasticenergies}) to express the sum of the activation energy times the corresponding electron impact rate outlined in terms of the reduced electric fields and reduced mobilities as follows:
\begin{align}
Q_{\rm e}
&= 
  \underbrace{\sum_k \beta_k^{\rm n} |C_{\rm e}| N_{\rm e} N_k (\mu_{\rm e}^\star)_k   \left( (E^\star_k)^2 -\frac{3  k_{\rm B}    (T_{\rm e}-T_{\rm ref})}{ m_{k} (\mu_{\rm e}^\star)_k^2}  \right)}_{\rm inelastic~losses~to~neutrals} \nonumber\alb
&+ \underbrace{ \sum_k  \frac{3 \beta_k^{\rm n} k_{\rm B} |C_{\rm e}| N_{\rm e} N_k  (T_{\rm e}-T)}{ m_k (\mu_{\rm e}^\star)_k} }_{\rm elastic~losses-gains~to~neutrals}       \nonumber\alb
 &+ \underbrace{\sum_k \beta_k^{\rm i}  N_{\rm e} N_k  (T_{\rm e}-T )  \frac{6 k_{\rm B} C_{\rm i}^2 C_{\rm e}^2 \ln \Lambda}{ \pi^3 \epsilon_0^2 m_{\rm e} m_k \overline{q_{\rm e}}^3}}_{\rm elastic~losses-gains~to~ions}  
 \label{eqn:Qe}
\end{align}
The latter includes all electron energy losses to the ions and the neutrals, either due to elastic or inelastic processes. It also includes electron energy gains due to elastic processes. However, it does not include electron energy gains due to inelastic processes.  In previous work, some have assumed that the rate of electron energy gain can be determined from the rate of electron energy loss by writing the loss-gain term as proportional to $T_{\rm e}-T_{\rm v}$ as in \cite{jap:2010:shneider} or \cite{pf:2021:parent}. \cite{jtht:2012:kim} have rather assumed that the rate of electron energy loss-gain due to inelastic collisions follows the Landau-Teller form. That is, it scales with the difference between the vibrational energy obtained from the electron temperature and the vibrational energy found from the gas temperature. However, such models were not derived from basic principles and were also not backed up by experimental data of electron heating. Therefore, we prefer to simply acknowledge that the rates of electron heating due to inelastic collisions are not known and to assume they are small and negligible even when the neutrals temperature is much higher than the electron temperature. As will be demonstrated in a subsequent section through comparisons with experimental data, in which the electron temperature is much less than the gas vibrational or translational temperature, such an assumption appears to be well justified.

\subsection{Reduced electric field for air neutral species}

\begin{figure*}[!h]
     \centering
     \subfigure[N$_2$]{~~~\includegraphics[width=0.3\textwidth]{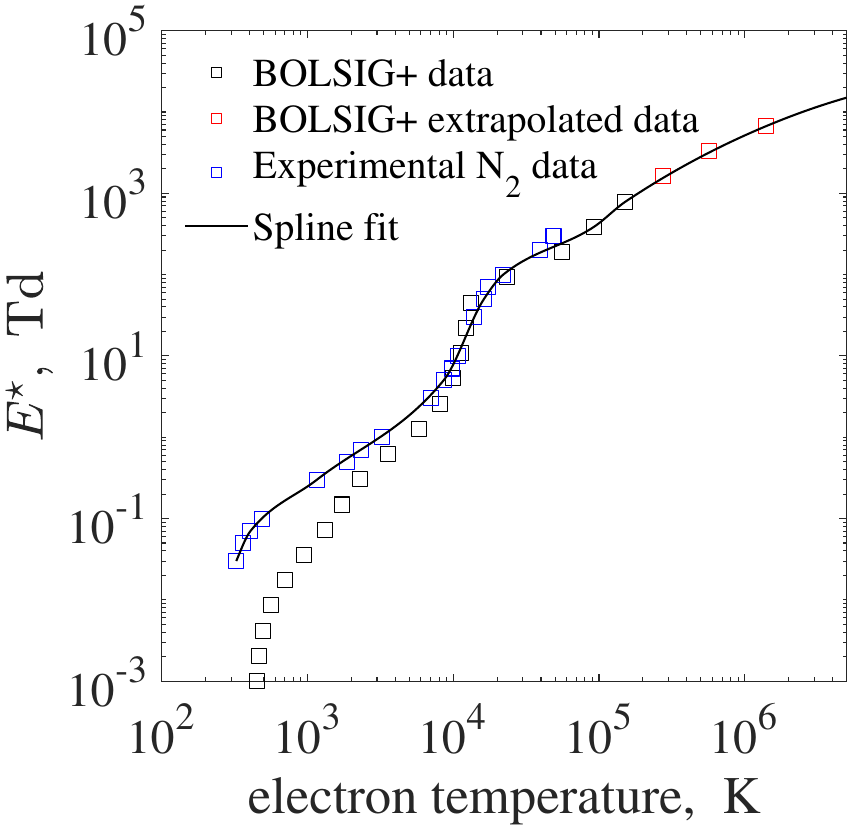}~~~}
     \subfigure[O$_2$]{~~~\includegraphics[width=0.3\textwidth]{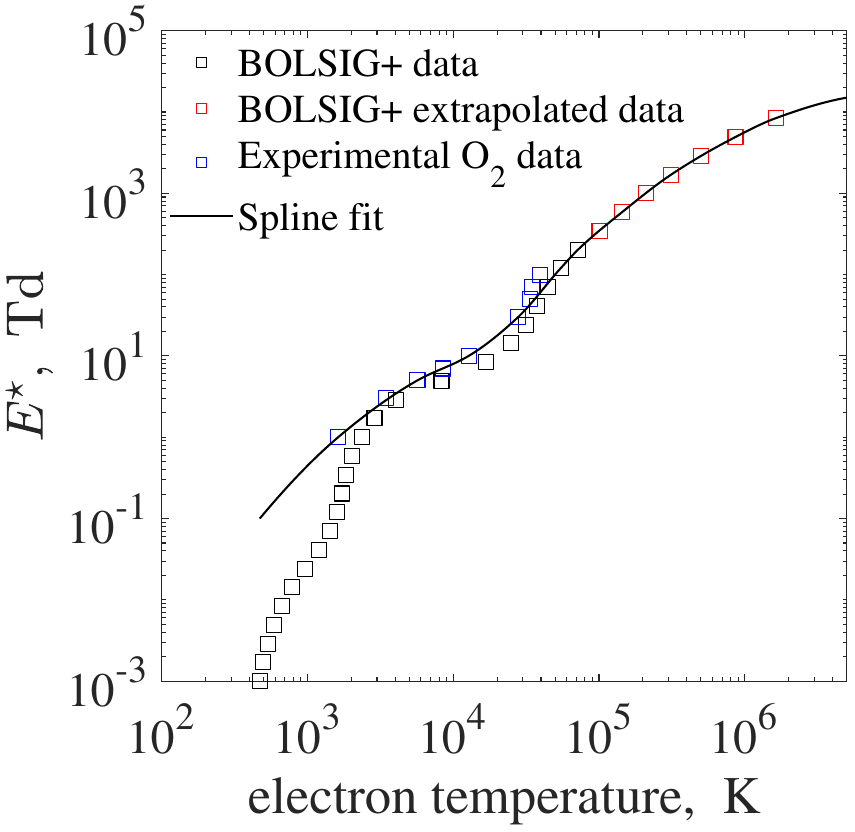}~~~}
    \subfigure[NO]{~~~\includegraphics[width=0.3\textwidth]{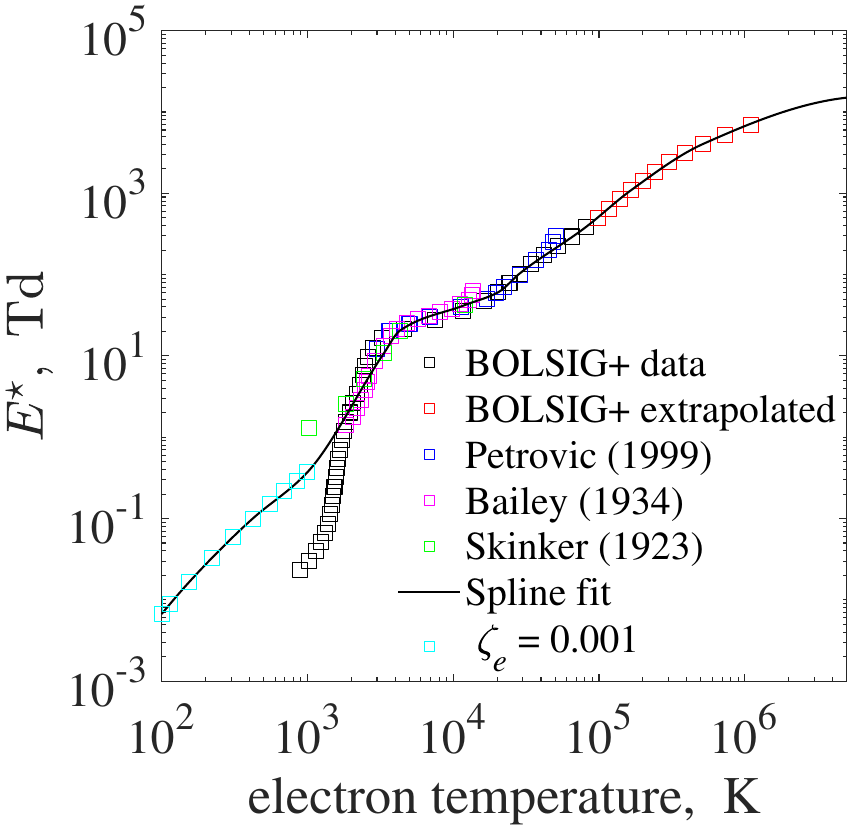}~~~}
        \subfigure[N]{~~~\includegraphics[width=0.3\textwidth]{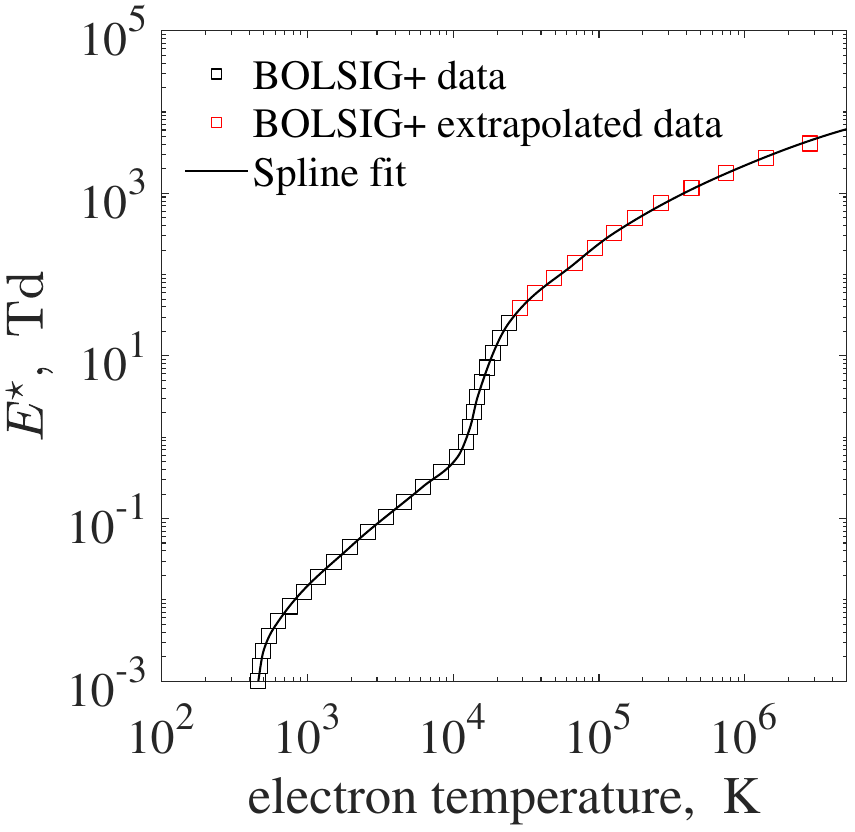}~~~}
        \subfigure[O]{~~~\includegraphics[width=0.3\textwidth]{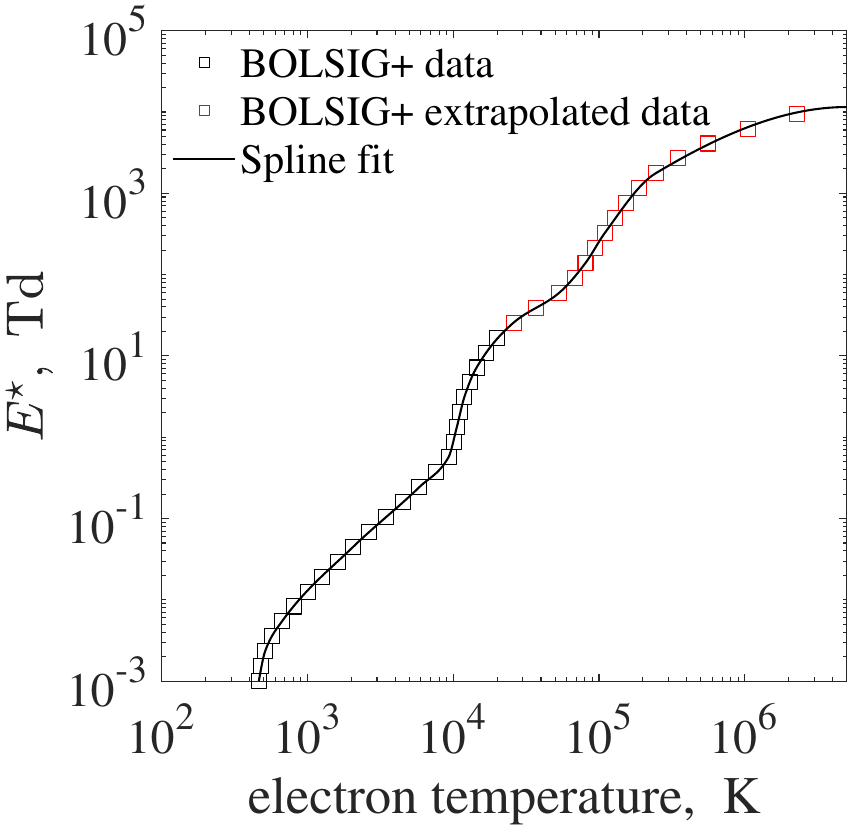}~~~}
     \figurecaption{Reduced electric field as a function of electron temperature for neutral species (a) $\rm N_2$, (b) $\rm O_2$, (c) $\rm NO$, (d) N, and (e) O.   }
     \label{fig:Estar_Te_splines}
\end{figure*}

\begin{table*}[!ht]
  \center\fontsizetable
  \begin{threeparttable}
    \tablecaption{Spline control points giving the reduced electric field from the electron temperature for air neutral species \tnote{a}.}
    \label{tab:EstarfromTe}
    \fontsizetable
 
    \begin{tabular*}{\textwidth}{@{}l@{\extracolsep{\fill}}ll@{}}
    
    \toprule
   Species~& ${\rm ln}~T_{\rm e}$ & ${\rm ln}~E^\star$  \\
        \midrule

   { N   }  &  \begin{minipage}[t]{0.4\textwidth}\raggedright  
 6.1322,    6.2124,    6.4474,    7.5872,    8.7256,    9.2774,    9.4778,    9.5934,   10.0915,   11.1390,   11.7497,   13.5255,   15.4249 \end{minipage}  & \begin{minipage}[t]{0.4\textwidth}\raggedright 
-55.2620,  -54.4162,  -53.5703,  -51.4558,  -49.7641,  -48.9182,  -48.0721,  -47.2264,  -45.1117,  -43.4205,  -42.5745,  -40.8824, -39.6301\end{minipage}\\
~\\
    { O   }  &  \begin{minipage}[t]{0.4\textwidth}\raggedright  
 6.1395,    6.2392,    6.5051,    7.6315,    8.6771,    9.1363,    9.2668,    9.3848,    9.5799,   10.1629,   10.8830,   11.2996,   11.6030,   12.4065, 15.4249 \end{minipage}  & \begin{minipage}[t]{0.4\textwidth}\raggedright 
 -55.2620,  -54.4162,  -53.5703, -51.4558,  -49.7641,  -48.9182,  -48.0721,  -47.2264,  -46.3806,  -45.1117,  -44.2660,  -43.4205,  -42.5745,  -40.8824,  -39.0042\end{minipage}\\
~\\
    {  N$_2$   }  &  \begin{minipage}[t]{0.4\textwidth}\raggedright  
    5.7836,    6.0067,    7.0566,    9.0580,    9.6956,   10.0010,   11.4313,   11.9302,   14.1582,   15.4249
  \end{minipage}  & \begin{minipage}[t]{0.4\textwidth}\raggedright 
   -51.8608,  -51.0135,  -49.5583,  -46.7448,  -44.4423,  -43.7491,  -42.3992,  -41.6848,  -39.5410, -38.7385
 \end{minipage}\\
 ~\\
    {  O$_2$  }     &  \begin{minipage}[t]{0.4\textwidth}\raggedright 
     6.1549,    7.3930,    8.6458,    9.4545,   10.2346,   10.9063,   11.1835,   11.8793,   12.6550,   13.6691,   14.3029,   15.4249
 \end{minipage}  & \begin{minipage}[t]{0.4\textwidth}\raggedright 
   -50.6569,  -48.3543,  -46.7448,  -46.0517,  -44.9531,  -43.5718, -43.0406,  -41.9779,  -40.9153,  -39.8524,  -39.3210,  -38.7385
 \end{minipage}\\  
 ~\\
     {  NO    }  &  \begin{minipage}[t]{0.4\textwidth}\raggedright 
  4.6052,    5.0387,    5.7384,    6.0438,    6.9078,    8.1163,    8.3684,    9.4004,    9.9245,   10.2589,   11.3016,   12.0194,   13.1593,   14.9141\end{minipage}  & \begin{minipage}[t]{0.4\textwidth}\raggedright 
 -53.3605,  -52.4569,  -51.1678,  -50.6572,  -49.3446,  -45.9734,  -45.3367,  -44.6079,  -44.2446,  -43.7196,  -42.4083,  -41.3594,  -40.0476,  -38.7385 \end{minipage}\alb

    \bottomrule
    \end{tabular*}
\begin{tablenotes}
\item[{a}] Notation and units: $T_{\rm e}$ is the electron temperature in Kelvin and $E^\star=|\vec{E}|/N$ is the reduced electric field in units of V$\rm ~m^2$.
\end{tablenotes}
   \end{threeparttable}
\end{table*}

Within the electron energy relaxation term $Q_{\rm e}$ shown in Eq.~(\ref{eqn:Qe}), the only terms that can not be readily determined are the reduced electron mobility $\mu_{\rm e}^\star$ and the reduced electric field $E^\star$. In this subsection, we will outline expressions for $E^\star$ for the commonly 5 neutral species within high temperature air: N$_2$, O$_2$, O, N, NO.

In obtaining a curve fit for the reduced electric field as a function of electron temperature, we utilize data from various sources, including experiments and BOLSIG+. The cross-section data required for BOLSIG+ is sourced from the Phelps database for the NO species and from the Morgan database for other species. Both the Morgan and Phelps cross-sectional data are extracted from the LXCat database by \cite{ppp:2017:pitchford}. For atomic species such as O and N, the processes encompass Townsend ionization, momentum transfer, and electronic excitation. Additionally, for molecular species, we include rotational and vibrational excitation processes. For either atomic or molecular species, electron attachment processes are excluded since they involve the loss of an electron and do not result in a loss of energy for the remaining electrons.

\begin{figure*}[!h]
     \centering
     \subfigure[$\rm N_2$]{~~~\includegraphics[width=0.3\textwidth]{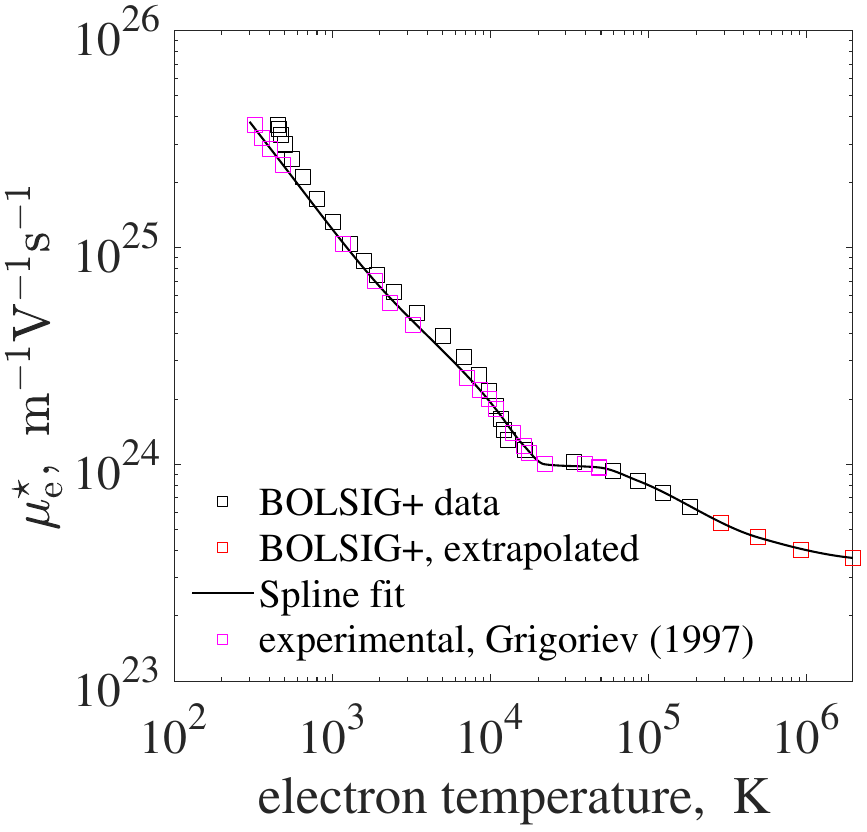}~~~}
     \subfigure[$\rm O_2$]{~~~\includegraphics[width=0.3\textwidth]{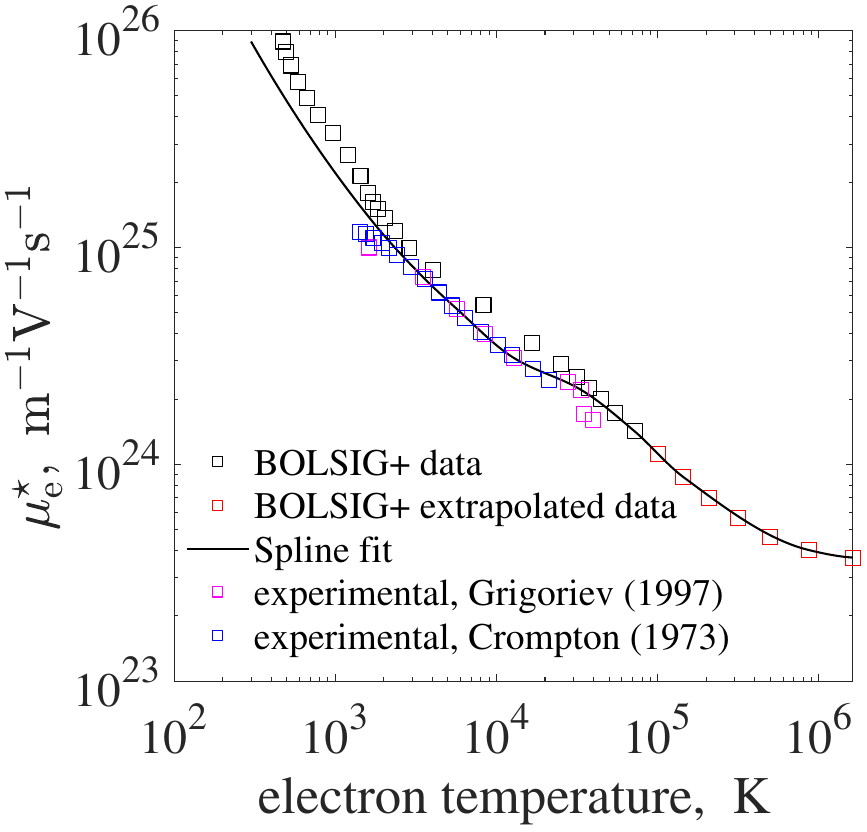}~~~}
     \subfigure[$\rm NO$]{~~~\includegraphics[width=0.3\textwidth]{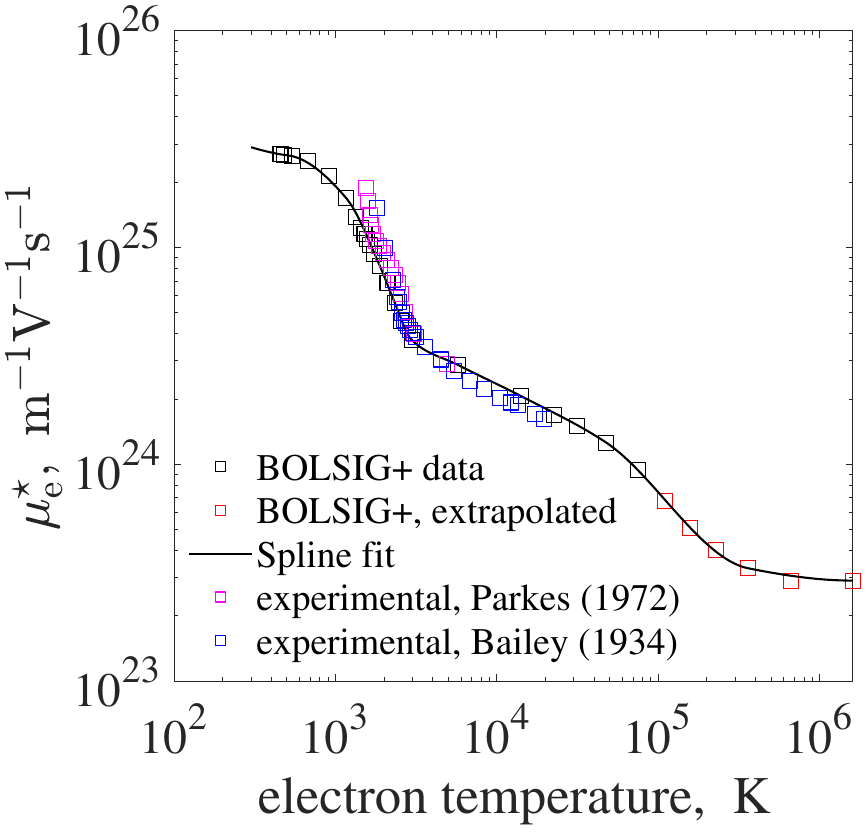}~~~}
     \subfigure[$\rm N$]{~~~\includegraphics[width=0.3\textwidth]{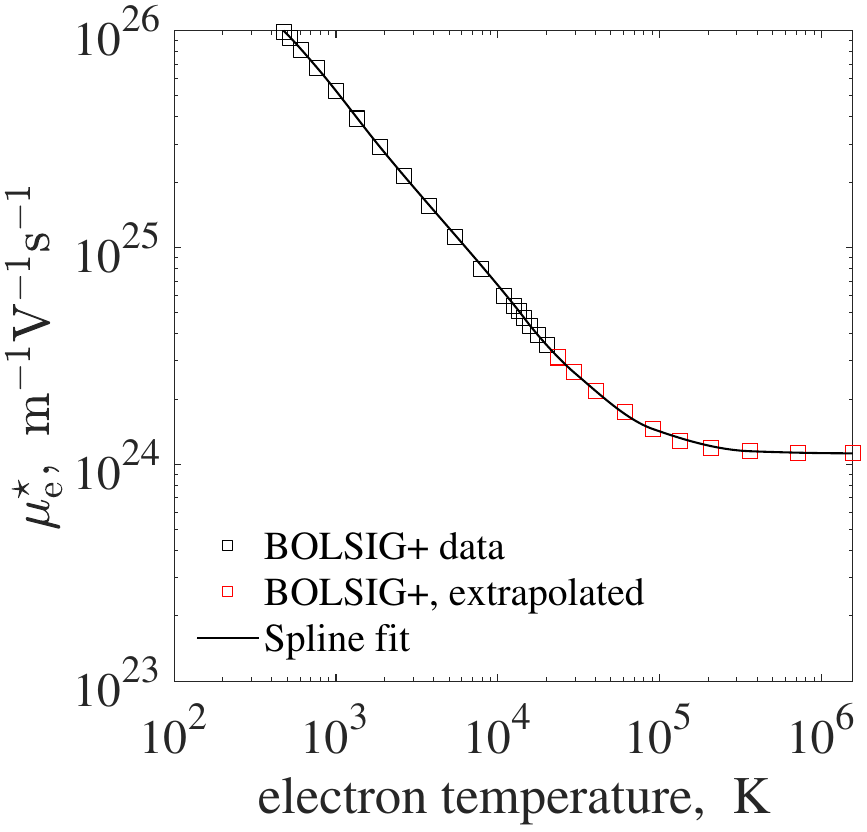}~~~}     
     \subfigure[$\rm O$]{~~~\includegraphics[width=0.3\textwidth]{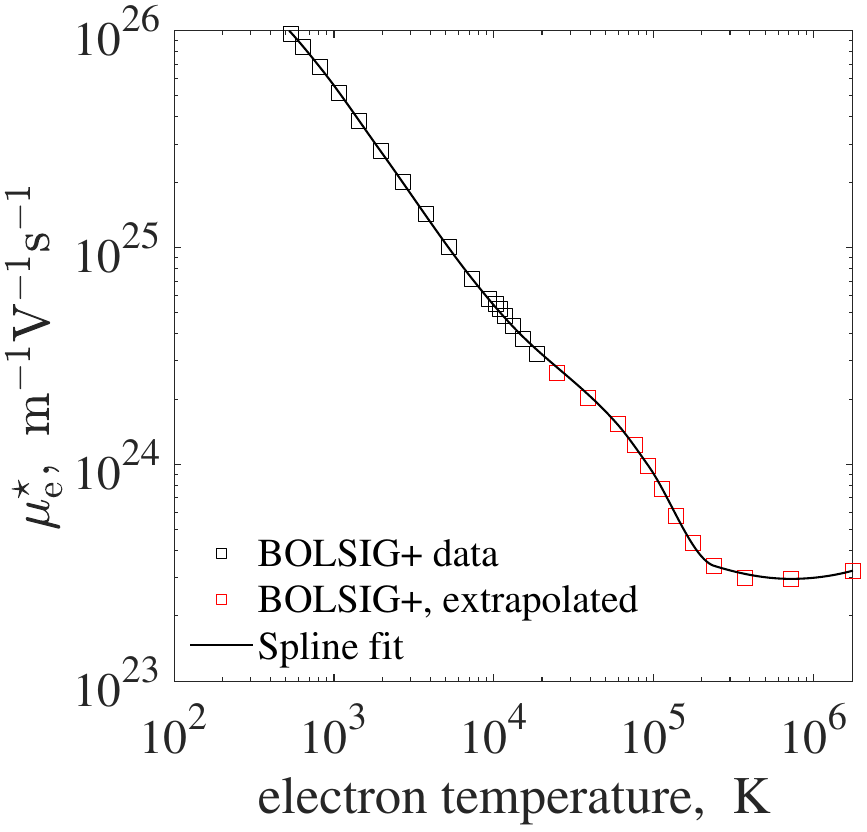}~~~}
     \figurecaption{Reduced electron mobility in neutral species as a function of electron temperature  (a) $\rm N_2$, (b) $\rm O_2$, (c) $\rm NO$, (d) N, and (e) O.   }
     \label{fig:mueN_Te_splines}
\end{figure*}

The experimental data for the reduced electric field as a function of electron temperature for both N$_2$ and O$_2$ is available from \cite[Ch.~21]{book:1997:grigoriev}. Experimental NO data points are taken from  \cite[Fig.~2]{jop:1999:mechlinska},  \cite[Fig.\ 5]{jos:1934:bailey} and \cite[Fig.\ 1.32]{jos:1934:skinker}. When the electron temperature is less than 1000~K, there is no experimental data for O$_2$ or NO. Because there are indications that the BOLSIG+ results at such low electron temperature are not accurate, we here rather extrapolate from the experimental data to lower $T_{\rm e}$. The extrapolation of the $\rm O_2$ curve for $T_{\rm e}<1000$~K is such that there is good agreement with known experimental data of air at low electron temperature, as will be shown in the section ``Test Cases'' below. The extrapolation of the NO curve for $T_{\rm e}<1000$~K  is such that the reduced electric field is computed from the following relation function of the electron energy loss function $\zeta_{\rm e} $ (the mean fraction of electron kinetic energy that the electron loses in one collision):
\begin{equation}
E^\star = \sqrt{\frac{3  k_{\rm B} T_{\rm e}\zeta_{\rm e} }{2 m_{\rm e} (\mu_{\rm e}^\star)^2}}       
\end{equation}
The latter relationship between the reduced electric field and the electron energy loss function can be easily derived starting from the electron energy transport in \cite[Eq.~(2.12)]{book:1991:raizer}, getting rid of terms function of spatial or temporal gradients,  and using the relationship between collision frequency and mobility outlined in Eq.~(\ref{eqn:collisionfrequency}). At a temperature of less than 1000~K, we here set $\zeta_{\rm e}$ to 0.001 because electrons in a molecular gas at low temperature mostly dissipate their energy by exciting the vibrational and rotational energy modes and such leads to a $\zeta_{\rm e}$ varying between $10^{-3}$ and $10^{-2}$ as outlined in \cite[page 17]{book:1991:raizer}. Such is one or two orders of magnitude higher than the dissipation of electron energy by elastic collisions ($\zeta_{\rm e}=2 m_{\rm e}/m_{\rm n} \approx 4\times10^{-5}$).

Monotone cubic splines by \cite{sjna:1980:fritsch} are fitted through the experimental data points (when available) and BOLSIG+ data (when no experimental data is available) for the 5 neutral species (see Fig.~\ref{fig:Estar_Te_splines}). The spline control points are listed in Table \ref{tab:EstarfromTe}.

\begin{table*}[!h]
  \center\fontsizetable
  \begin{threeparttable}
    \tablecaption{Spline control points giving the reduced electron mobility from the electron temperature for air neutral species \tnote{a}.}
    \label{tab:mueNfromTe}
    \fontsizetable
 
    \begin{tabular*}{\textwidth}{@{}l@{\extracolsep{\fill}}ll@{}}
    
    \toprule
   Species~& ${\rm ln}~T_{\rm e}$ & ${\rm ln}~\mu_{\rm e}^\star$  \\
        \midrule

   { N   }  &  \begin{minipage}[t]{0.4\textwidth}\raggedright  
     5.7038,    6.1322,    6.6388,    7.5322,    9.5102,    9.6725,    9.9096,   10.0721,   10.2909,   11.4162,   12.8047,   14.2629
 \end{minipage}  & \begin{minipage}[t]{0.4\textwidth}\raggedright 
   60.1673,   59.8890,   59.4736,   58.6331,   56.8895,   56.7320,   56.5248,   56.3967,   56.2404,   55.6370,  55.4009,   55.3789
\end{minipage}\\
~\\
   { O   }  &  \begin{minipage}[t]{0.4\textwidth}\raggedright  
        5.7038,    6.1395,    6.7054,    7.5795,    8.8973,    9.4805,   10.9925,   11.4348,   12.3840,   13.4898,   14.3824
 \end{minipage}  & \begin{minipage}[t]{0.4\textwidth}\raggedright 
     60.2388,   59.9607,   59.4786,  58.5874,   57.2324,   56.7285,   55.6853,   55.2478,   54.1838,   54.0457,   54.1329
\end{minipage}\\
~\\
   { $\rm N_2$   }  &  \begin{minipage}[t]{0.4\textwidth}\raggedright  
           5.7038,    6.0067,    7.5266,    9.0580,    9.2866,    9.6956,   10.0010,   10.7942,   11.3595,  13.1075,   14.4873
 \end{minipage}  & \begin{minipage}[t]{0.4\textwidth}\raggedright 
   58.8996,   58.6144,   57.2080,   56.0505,   55.8498,   55.4609,   55.2620,   55.2282,   55.0865,   54.4855,   54.2678
\end{minipage}\\
~\\
   { $\rm O_2$   }  &  \begin{minipage}[t]{0.4\textwidth}\raggedright  
    5.7038,    7.4529,    8.1552,    8.6458,    9.4545,   10.4239,   11.1835,   11.8793,   13.6691,   14.3029
 \end{minipage}  & \begin{minipage}[t]{0.4\textwidth}\raggedright 
     59.7518,   57.8270,   57.2544,   56.9107,   56.3934,   56.0505,   55.6127,   55.1276,   54.3525,   54.2716
\end{minipage}\\
~\\
   { $\rm NO$   }  &  \begin{minipage}[t]{0.4\textwidth}\raggedright  
      5.7038,    6.1119,    6.2863,    7.0606,    7.5408,    7.7537,    7.9923,    8.6608,    9.5506,   10.7689,   12.7980,   14.2900
 \end{minipage}  & \begin{minipage}[t]{0.4\textwidth}\raggedright 
    58.6293,   58.5553,   58.5354,   58.0894,   57.3686,   56.9769,   56.5835,   56.3153,   55.9881,   55.4932,   54.1576,   54.0273
\end{minipage}\alb
    \bottomrule
    \end{tabular*}
\begin{tablenotes}
\item[{a}] Notation and units: $T_{\rm e}$ is the electron temperature in Kelvin and $\mu_{\rm e}^\star=\mu_{\rm e}N$ is the reduced electron mobility in units of $\rm V^{-1}~m^{-1}~s^{-1}$.
\end{tablenotes}
   \end{threeparttable}
\end{table*}

\subsection{Reduced mobilities for air neutral species}

In this subsection, expressions are outlined for the reduced mobility of electrons within the neutral species. We will do so for molecular nitrogen, molecular oxygen, nitrogen oxide and atomic oxygen and nitrogen.

In obtaining a curve fit for the reduced electron mobility as a function of electron temperature, we utilize data from experiments and BOLSIG+. We use the same cross-sectional data to determine the reduced mobility as the one we used in the previous subsection to determine the reduced electric field. Thus, the cross-section data for NO is sourced from the Phelps database in \cite{jcp:2012:pancheshnyi}, while for other species, it is sourced from the \cite{pcpp:1992:morgan} database. The cross-sections are limited to Townsend ionization, momentum transfer, and electronic excitation for the atomic species, and further include rotational and vibrational excitation processes for the molecular species.

For molecular nitrogen and oxygen, experimental data of the reduced mobility is obtained  from  \cite[Ch.\ 21]{book:1997:grigoriev}. A second source of experiments for molecular oxygen is found in \cite[Fig.~4]{ajp:1973:crompton}. As for nitric oxide experimental data is obtained from  \cite[Fig.\ 4]{jos:1934:bailey} and  from \cite[Fig.\ 5]{jcs:1972:parkes}. When the experimental data is given in terms of drift velocity rather than reduced mobility, we convert it to reduced mobility using the relationship $\mu_{\rm e}^\star=v_{\rm drift}/E^\star$ with the reduced electric field taken from the spline fits shown in the previous subsection.

The \cite{sjna:1980:fritsch} monotone cubic splines are fitted through experimental data when available and through BOLSIG+ data when experiments are lacking. The various data as well as the spline curve fits are shown in Fig.~\ref{fig:mueN_Te_splines}, while the spline control points for the reduced mobility of all air neutral species are outlined in Table \ref{tab:mueNfromTe}.

\begin{table*}[!h]
\fontsizetable
\begin{center}
\begin{threeparttable}
\tablecaption{Corrected air plasma reaction rates.}
\begin{tabular*}{\textwidth}{@{}l@{\extracolsep{\fill}}lll@{}} 
\toprule

{Reaction} &  Rate coefficient & {Ref.}\\ 
\midrule                

 $\rm  e^- +  NO^{+}  \rightarrow  N + O $   & $3.00 \cdot 10^{-7}\cdot(T_{\rm e}/300)^{-0.56}~~{\rm cm^3/s}$ & \cite{jgr:2004:sheehan} \\

 $\rm  e^- + O_2^{+} \rightarrow  O + O $  & $2.40 \cdot 10^{-7}\cdot(T_{\rm e}/300)^{-0.70}~~{\rm cm^3/s}$ & \cite{aip:2001:peverall} \\

$\rm e^- + N_2^{+}\rightarrow N + N $   &  Rate given as spline-fit through rate data   & \cite[Fig.~3]{aip:1998:peterson} \\

$\rm e^{-} +N_2\rightarrow N + N + e^{-}$ & Rate given as spline-fit using cross-section data    & \cite[Fig.~10]{pr:1985:phelps} \\

  $\rm e^- + e^- + N^{+}\rightarrow e^- + N $ & $2.20 \cdot 10^{40} \cdot{\cal A}^{-2}\cdot T_{\rm e}^{-4.5}~~{\rm cm^6/s}$
                                               &\cite{nasa:1973:dunn}\\

  $\rm e^- + e^- + O^{+} \rightarrow e^- + O $ & $2.20 \cdot 10^{40} \cdot{\cal A}^{-2}\cdot T_{\rm e}^{-4.5}~~{\rm cm^6/s}$
                                               &\cite{nasa:1973:dunn}\\

 $\rm e^- + N \rightarrow N^{+}+e^- + e^-$ & Rate given as spline-fit using cross-section data    &\cite[Fig.~4]{pr:1962:smith}\\

 $\rm e^- + O \rightarrow O^{+}+e^- + e^-$  & Rate given as spline-fit using cross-section data &\cite{pcpp:1992:morgan}\\

 $\rm e^- + N_2 \rightarrow N_{2}^{+}+e^- + e^-$ &Rate given as spline-fit using cross-section data   &\cite{pcpp:1992:morgan}\\
 $\rm e^- + O_2 \rightarrow O_{2}^{+}+e^- + e^-$ & Rate given as spline-fit using cross-section data  &\cite{pcpp:1992:morgan}\\  

 $\rm e^- + NO \rightarrow NO^{+}+e^- + e^-$ & Rate given as spline-fit using cross-section data  &\cite{pcpp:1992:morgan}\\  
                                                         
\bottomrule
\end{tabular*}
\begin{tablenotes}
\item[{a}] In the rate expressions, $T_{\rm e}$ is the electron temperature in Kelvin. ${\cal A}$ is Avogadro's number and it is approximately $6.022 \cdot 10^{23}~{\rm mol^{-1}}$.
\item[{b}] The spline control points for each reaction can be found in Table \ref{tab:11s_spline_tab}. The corresponding reaction rates (in ${\rm cm^3/s}$) are obtained with BOLSIG+ using cross-section data from the indicated references or by directly fitting a cubic spline through experimentally measured reaction rate values.
\end{tablenotes}
\label{tab:correctedreactionrates}
\end{threeparttable}
\end{center}
\end{table*}

\begin{figure*}[!h]
     \centering
     \subfigure[$\rm e^- + N \rightarrow N^{+}+e^- + e^-$]{~~~\includegraphics[width=0.29\textwidth]{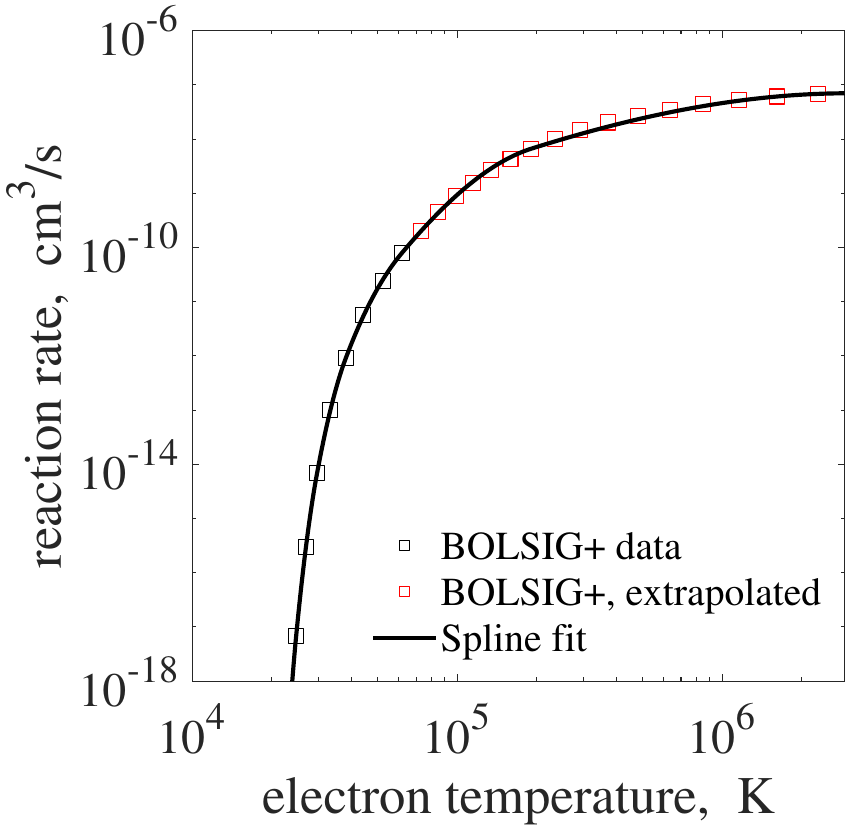}~~~}
     \subfigure[$\rm e^- + O \rightarrow O^{+}+e^- + e^-$]{~~~\includegraphics[width=0.29\textwidth]{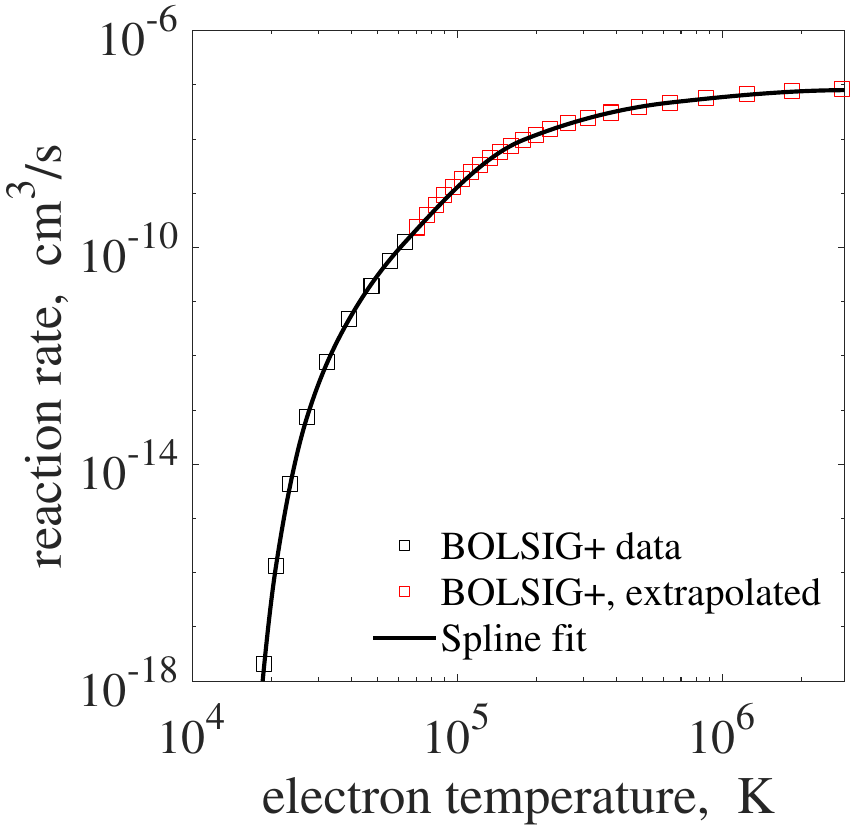}~~~}
     \subfigure[$\rm e^- + N_{2}^+ \rightarrow N+N$]{~~~\includegraphics[width=0.29\textwidth]{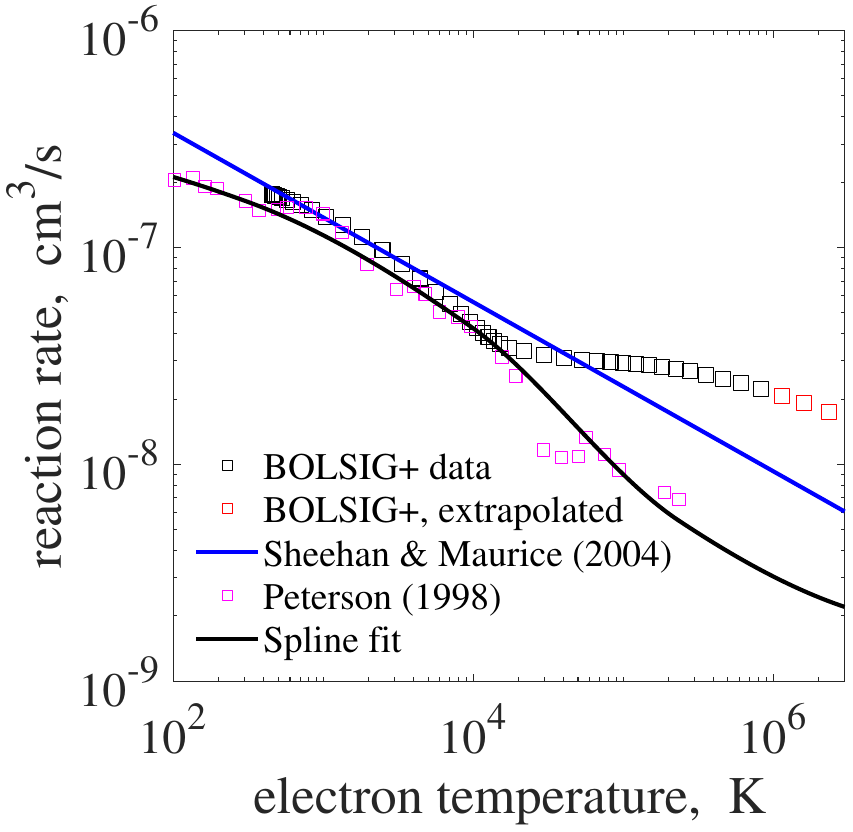}~~~}
     \subfigure[$\rm e^- + N_{2} \rightarrow N+N + e^-$]{~~~\includegraphics[width=0.29\textwidth]{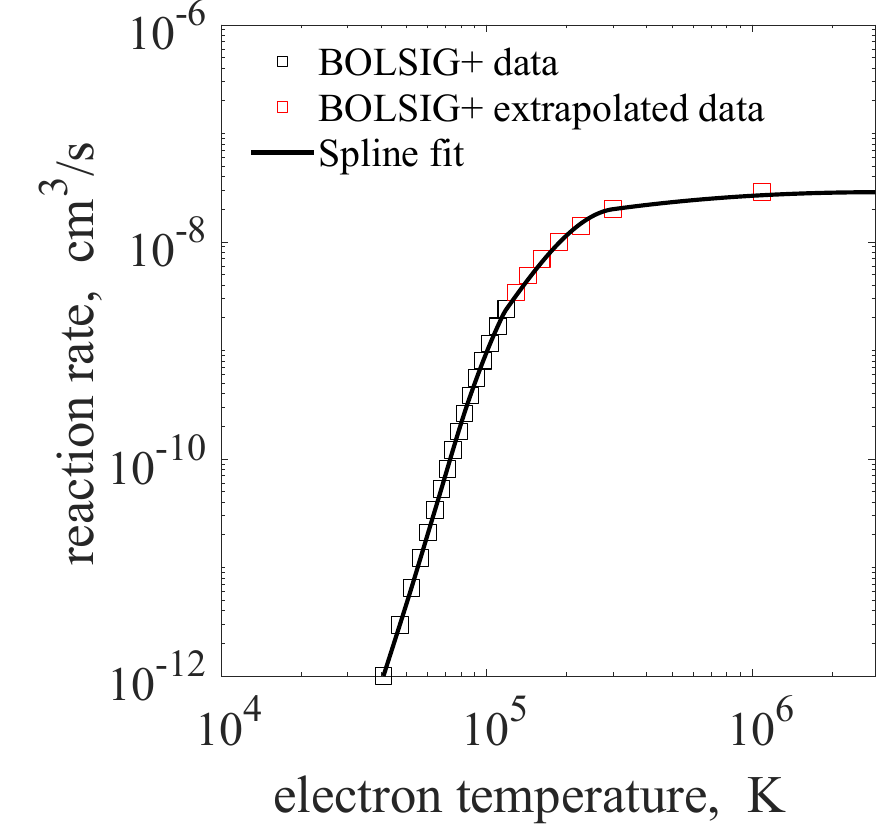}~~~}
     \subfigure[$\rm e^- + N_{2} \rightarrow N_{2}^+ +  e^-  + e^-$]{~~~\includegraphics[width=0.29\textwidth]{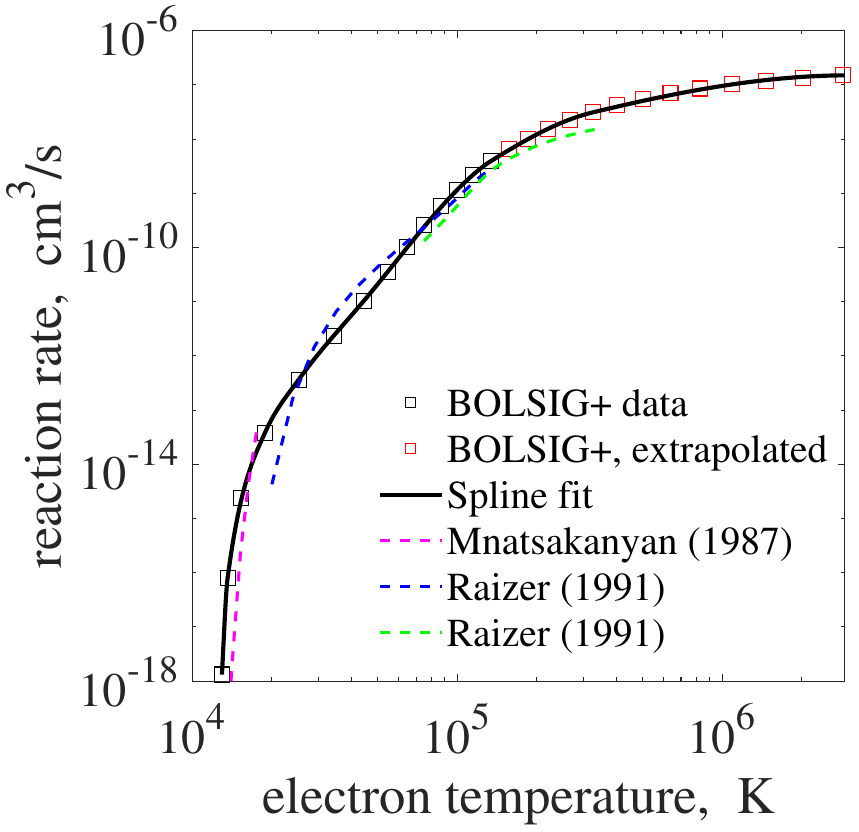}~~~}
     \subfigure[$\rm e^- + O_{2} \rightarrow O_{2}^+ +  e^-  + e^-$]{~~~\includegraphics[width=0.29\textwidth]{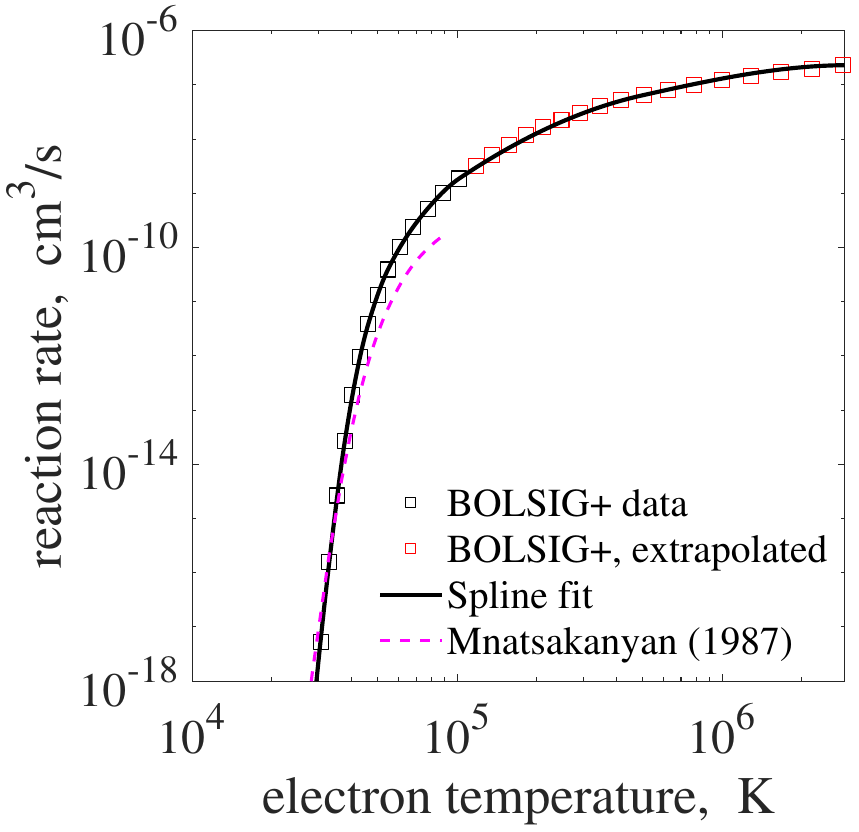}~~~}
     \subfigure[$\rm e^- + NO \rightarrow NO^{+}+e^- + e^-$]{~~~\includegraphics[width=0.29\textwidth]{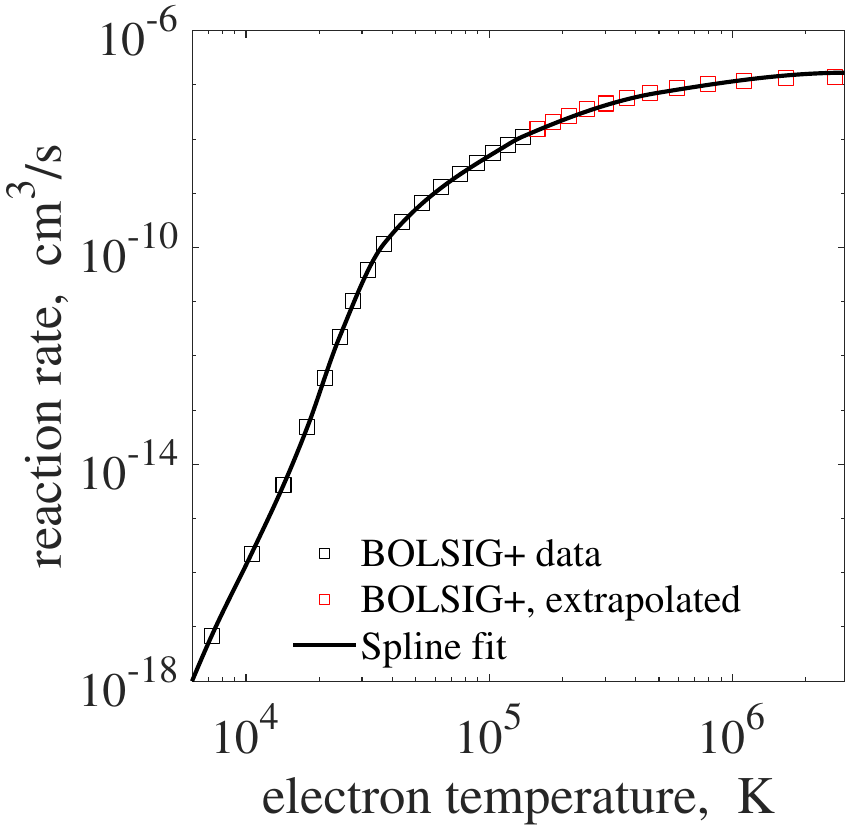}~~~}
     \figurecaption{Corrected air plasma reaction rates function of electron temperature.}
     \label{fig:reactionrate_Te_splines}
\end{figure*}

\begin{table*}[!ht]
  \center\fontsizetable
  \begin{threeparttable}
    \tablecaption{Coordinates of spline control points for the corrected air plasma reaction rates.\tnote{a}}
    \label{tab:11s_spline_tab}
    \fontsizetable
 
    \begin{tabular*}{\textwidth}{@{}l@{\extracolsep{\fill}}llll@{}}
    
    \toprule
    Reaction ~ & ${\rm ln}~T_{\rm e}$ & ${\rm ln}~k$ \\
        \midrule
        
     { $\rm e^- + N \rightarrow N^+ + e^- +e^- $   } &   \begin{minipage}[t]{0.30\textwidth}\raggedright  
          9.9733,   10.1148,   10.2972,   10.5461,   11.0331,   12.1595,   14.9141

 \end{minipage}  & \begin{minipage}[t]{0.35\textwidth}\raggedright 
    -49.6819,  -39.5251,  -32.5922,  -27.7250,  -23.2627,  -18.8262,  -16.4748

\end{minipage} \\       
~\\

 { $\rm e^- +O \rightarrow O^+ + e^- +e^- $   } &   \begin{minipage}[t]{0.3\textwidth}\raggedright  
    9.7444,    9.9402,   10.2102,   10.5776,   11.0606,   12.0831,   13.3654,   14.9141

 \end{minipage}  & \begin{minipage}[t]{0.35\textwidth}\raggedright 
  -45.7081,  -36.5368,  -30.2192,  -26.0516, -22.8011,  -18.4691, -16.8693,  -16.3412

\end{minipage} \\      
~\\

   { $\rm e^- +  N_2^+ \rightarrow   N+N  $   } &  \begin{minipage}[t]{0.3\textwidth}\raggedright  
            2.9613,    5.7153,    9.1741,   12.3696,   14.9141

 \end{minipage}  & \begin{minipage}[t]{0.3\textwidth}\raggedright 
   -15.1097,  -15.6251,  -16.9585,  -19.0192,  -19.9348

\end{minipage} \\
~\\

   { $\rm e^- + N_2 \rightarrow N+N + e^-$   } & \begin{minipage}[t]{0.3\textwidth}\raggedright  
    9.5662, 9.6421, 9.8483, 9.9808, 10.2742, 10.8312, 10.9937, 11.1449, 11.4428, 11.6894, 14.9141
 \end{minipage}  & \begin{minipage}[t]{0.35\textwidth}\raggedright 
  -45.6975, -41.8099, -35.8672, -33.6513, -30.3852, -26.0004, -24.6913, -23.4619, -21.1866, -19.8272, -13.9209
\end{minipage} \\
~\\

   { $\rm e^- + N_2 \rightarrow N_{2}^+ +e^- + e^-$   } & \begin{minipage}[t]{0.3\textwidth}\raggedright  
         9.4698,    9.5196,    9.8428,   10.1366,   10.7023,   11.6533,   11.8041,   12.6903,   14.9141

 \end{minipage}  & \begin{minipage}[t]{0.35\textwidth}\raggedright 
     -41.1651,  -37.0451,  -30.9051,  -28.6392,  -25.2960,  -19.9307,  -19.3649,  -17.2893,  -15.7126

\end{minipage} \\
~\\

   { $\rm e^- + O_2 \rightarrow O_{2}^+ +e^- + e^-$   } & \begin{minipage}[t]{0.3\textwidth}\raggedright  
    10.2400,   10.4030,   10.5381,   10.6702,   10.7424,   10.9127,   11.2568,   11.5315,   13.1215,   14.9141

 \end{minipage}  & \begin{minipage}[t]{0.35\textwidth}\raggedright 
   -44.0087,  -36.3676,  -31.2400,  -27.6638,  -26.2591,  -23.9601,  -21.3797,  -20.0957,  -16.5644,  -15.2808,

\end{minipage} \\
~\\

   { $\rm e^- + NO \rightarrow NO^+ +e^- + e^-$   } & \begin{minipage}[t]{0.3\textwidth}\raggedright  
         8.4874,    8.8842,    9.2631,    9.5654,    9.7903,   10.0982,   10.5173,   11.6903,   11.8306,   13.1133,   14.9141

 \end{minipage}  & \begin{minipage}[t]{0.35\textwidth}\raggedright 
   -43.8225  -39.5270  -36.0264  -33.1178  -30.6688  -26.8095  -22.8654  -18.6549  -18.3163  -16.4545  -15.5993

\end{minipage} \alb
 
    \bottomrule
    \end{tabular*}
\begin{tablenotes}
\item[{a}] The reaction rate $k$ has units of $\textrm{cm}^3\cdot \textrm{s}^{-1}$. The electron temperature $T_{\rm e}$ is in Kelvin.

\end{tablenotes}
\label{tab:correctedreactionratessplinecontrolpoints}
   \end{threeparttable}
\end{table*}

\section{Air Plasma Chemical Reactions}

When assessing the accuracy of the proposed electron temperature equation through comparison with experimental results, we will utilize either a low-temperature 8-species air chemical model incorporating negative ions by \cite{jpp:2007:parent} or a Park-like 11-species high-temperature air chemical model by \cite{ijhmt:2021:kim}. However, we adjust the reaction rates of several chemical reactions involving electrons in the reactants. This modification is necessary to minimize the physical errors arising from the chemical solver.

The corrected reactions for both the low-temperature and high-temperature air chemical solvers are displayed in Table \ref{tab:correctedreactionrates}, with the control points of spline fits listed in Table \ref{tab:correctedreactionratessplinecontrolpoints}. The \cite{sjna:1980:fritsch} monotone cubic spline curve fits, along with the BOLSIG+ data and experimental data upon which they are based, are illustrated in Fig.~\ref{fig:reactionrate_Te_splines}.

When obtaining the Townsend ionization rates from BOLSIG+, we utilize cross sections sourced from the \cite{pcpp:1992:morgan} database  for O, $\rm N_2$, and $\rm O_2$, from the Phelps database in \cite{jcp:2012:pancheshnyi} for NO, and ionization cross sections from \cite{pr:1962:smith} for N. For the dissociative recombination of NO$^+$, we use cross-sectional data listed in  \cite{aip:1998:peterson}.

The experimental data curve fits for the Townsend ionization rate of $\rm N_2$ is taken from \cite{kp:1987:mnatsakanyan} for the range $3\cdot10^{-20}<E^\star<3\cdot10^{-19}~~\rm V~m^2$, and from \cite[pages 55-56]{book:1991:raizer} for a higher range valid up to $2.4\cdot10^{-18}~~\rm V~m^2$.  For the Townsend ionization of $\rm O_2$ and for the dissociative recombination rate of $\rm N_{2}^+$, the experimental data are taken from \cite{kp:1987:mnatsakanyan} and from \cite{aip:1998:peterson}, respectively. Wherever the experimental data is given as a function of the reduced electric field $E^\star$ rather than the electron temperature $T_{\rm e}$, we convert the rates using the relationship between the reduced electric field and the electron temperature outlined in Table \ref{tab:EstarfromTe}. 

When a close agreement between experiments and BOLSIG+ is observed, the \cite{sjna:1980:fritsch} monotone cubic splines   are fitted to the BOLSIG+ data. This is the case for all reactions except for the dissociative recombination rate of $\rm N_{2}^+$, where a significant disagreement is apparent (see Fig.~\ref{fig:reactionrate_Te_splines}c). For this reaction, we fit a cubic spline through experimental data from \cite{aip:1998:peterson} rather than through the BOLSIG+ values. Here, the BOLSIG+ predictions seem to deviate considerably at electron temperatures higher than 20,000 K, not only in comparison to experiments but also in comparison to a rate estimate in Arrhenius form from \cite{jgr:2004:sheehan}.

\section{Numerical Methods}

The mass, momentum, and energy transport equations outlined in Section 2 above, including the electric field potential equation, are first recast to eliminate the stiffness associated with solving Gauss's law following the method of \cite{book:2022:parent}. It is emphasized that this recasting does not alter the physical model in any way: no terms are discarded, and no new terms are added. Instead, the transport equations are rearranged such that the Gauss-based potential equation becomes an Ohm-based potential equation, thus avoiding the extremely stiff terms on the right-hand side of the Gauss-based potential equation. As demonstrated in \cite{jcp:2014:parent}, this recasting guarantees satisfaction of Gauss's law and results in the same converged solution as when using the standard transport equations.

The recast set of mass, momentum, and energy transport equations is discretized using the \cite{jcp:1981:roe} scheme, made second-order accurate through the Monotonic Upstream-centered Scheme for Conservation Laws (MUSCL) scheme, and the \cite{jcp:1979:vanleer} Total Variation Diminishing limiter. For the RAM-C-II cases, a positivity-preserving limiter by \cite{aiaaconf:2019:parent}, as well as an entropy correction based on the Peclet number by \cite{aiaa:2017:parent}, are applied to the discretized flux for fast and reliable convergence.

The discretized fluid transport equations are iterated to steady-state using a diagonally-dominant alternate direction block-implicit method (DDADI) by \cite{aiaaconf:1987:bardina}, while the potential equation is solved using a combination of the iterative modified approximate factorization (IMAF) scheme by \cite{cf:2001:maccormack} and the successive over relaxation (SOR) scheme of \cite{gen:douglas}. 

The physical model outlined in Section 2 along with the discretization stencils and the integration algorithms outlined in this section are implemented in the in-house-developed code CFDWARP (Computational Fluid Dynamics, Waves, Reactions, Plasmas). Such is the code used to obtain all results shown herein.

\section{Test Cases}

The proposed electron energy transport model's accuracy is now assessed through three different test cases: a uniform plasma subjected to a uniform electric field, the RAM-C-II re-entry flight test, and a glow discharge within a hypersonic boundary layer. Each case will be compared with experimental data, and where feasible, evaluated against other models of electron energy transport. 

\subsection{Uniformly Applied Electric Field on Plasma Flow}

\begin{figure}[!b]
     \centering
     \includegraphics[width=0.36\textwidth]{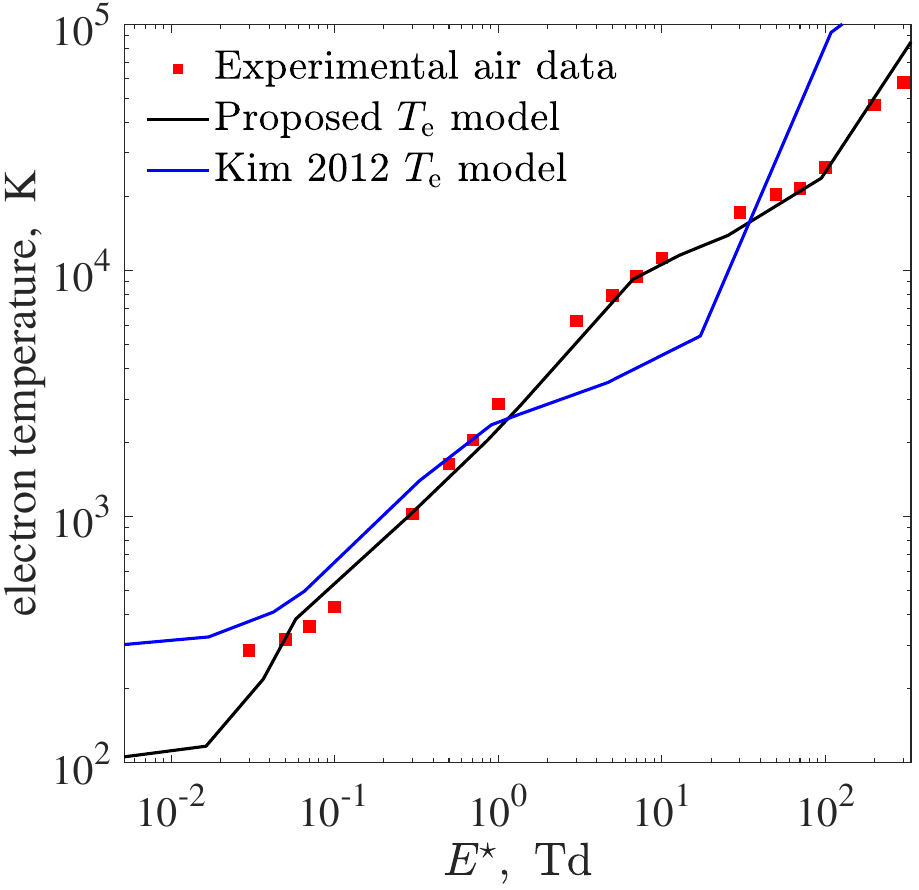}
     \figurecaption{Electron temperature as a function of the reduced electric field for the uniformly applied electric field on a plasma flow test case.}
     \label{fig:appliedE_Te_vs_Estar}
\end{figure}

The first test case involves applying a uniform electric field to a pre-ionized air plasma flow. The plasma enters the computational domain with a velocity of 2000 m/s, a pressure of 1 atm, a temperature of 300 K, an electron molar fraction of $10^{-10}$, and a $\rm N_2^+$ molar fraction of also $10^{-10}$. The 2D computational domain measures 3 cm in length and 1 cm in height. Boundary conditions for the domain include a supersonic inflow on the left, supersonic outflow on the right, and symmetry conditions on the top and bottom. We apply uniform energy deposition $\vec{E}\cdot \vec{J}_{\rm e}= \vec{E}\cdot \vec{J}$ across the entire domain to induce an increase in electron temperature. The energy deposition varies between $10^2$ to $10^9$ W/m$^3$. It is ensured that this does not significantly affect gas temperature, pressure, or number density. To maintain constant electron and ion densities throughout the domain, chemical reactions are disabled. Additionally, due to the low gas temperature, the \cite{book:1984:dixon-lewis} model  is utilized to determine the transport coefficients, with mobilities as specified in Table~\ref{tab:mueNfromTe}.  Because the region where the electron temperature and other properties are measured is free of significant gradients, the solution shows little sensitivity to the grid and a $10\times 5$ mesh is deemed sufficient for minimal numerical error. 

Noting that for a uniform plasma the component of the current related to the pressure gradients becomes zero, the electron current becomes simply $\vec{J}_{\rm e}=\mu_{\rm e} N_{\rm e} |C_{\rm e}| \vec{E} $.  Then, the energy input to the electrons can be written as:
\begin{equation}
\vec{E}\cdot\vec{J}_{\rm e} = \mu_{\rm e} N_{\rm e} |C_{\rm e}| \vec{E}^2
\end{equation}
We can then multiply and divide the right-hand-side by $N^2$, and isolate the reduced electric field $E^\star=|\vec{E}|/N$ to obtain:
\begin{equation}
E^\star=\frac{1}{N}\sqrt{\frac{\vec{E}\cdot\vec{J}_{\rm e}}{\mu_{\rm e} N_{\rm e} |C_{\rm e}|}} 
\label{eqn:EstarfromEdotJe}
\end{equation}
All the properties on the right-hand-side are either specified in the control file or can be measured within the flow. 

We proceed as follows: for a specified $\vec{E}\cdot\vec{J}_{\rm e}$, we obtain a converged solution and then measure at a certain location near the outflow boundary the electron temperature as well as the electron mobility, the total number density, and the electron number density. By varying $\vec{E}\cdot\vec{J}_{\rm e}$ over a large range of $10^2$ to $10^9$ W/m$^3$, we can then obtain electron temperature as a function of the reduced electric field over an electron temperature range spanning from 300~K to 60,000~K. This data can be compared to experimental data of air at sea-level conditions as outlined in \cite[Ch.~21]{book:1997:grigoriev}. 

In Fig.~\ref{fig:appliedE_Te_vs_Estar}, a comparison is presented between the proposed electron temperature model, the \cite{jtht:2012:kim} electron temperature model, and experimental data from \cite[Ch.~21]{book:1997:grigoriev}. It is  emphasized that the experiments are conducted under identical conditions of gas composition, temperature, and pressure as those utilized in the simulations. The proposed model yields a solution that closely matches experimental values (with an error not exceeding 25\%) across the entire range of reduced electric field where experimental data is available. In contrast, previous models of electron temperature for hypersonic flows (such as the Kim 2012 model) exhibit much greater discrepancy with the experimental findings. Specifically, the electron temperature obtained with the Kim 2012 model can be as much as  2-3 times lower or higher than the values obtained through experiments.

Particular care has been taken to ensure the correct implementation of the Kim 2012 model. One key difference between the Kim 2012 model and the one proposed herein is that it requires the input of electron energy losses through inelastic electron-impact collisions. This requirement is not present in the proposed model because all such losses are already incorporated within the reduced electric field terms. To adapt the Kim 2012 model, we include the electron-impact losses due to Townsend ionization by multiplying the Townsend ionization rates listed in Table \ref{tab:correctedreactionrates} by their respective ionization potentials. For species N, O, N$_2$, O$_2$, and NO, the ionization potentials are set to 14.54, 13.61, 15.65, 12.15, and 9.39 eV, respectively. However, it is noteworthy that adding the electron energy loss due to Townsend ionization does not significantly affect the results across the entire electron temperature range considered here ($T_{\rm e}<60,000$~K). Instead, the low accuracy of the Kim 2012 model is attributed to errors in the modeling of electron energy losses to the vibrational energy modes of the nitrogen molecule because such losses dominate at the relatively low electron temperatures encountered in this test case.

\subsection{RAM-C-II Flight Test}

\begin{figure*}[!t]
     \centering
     \subfigure[]{~~~\includegraphics[width=0.32\textwidth]{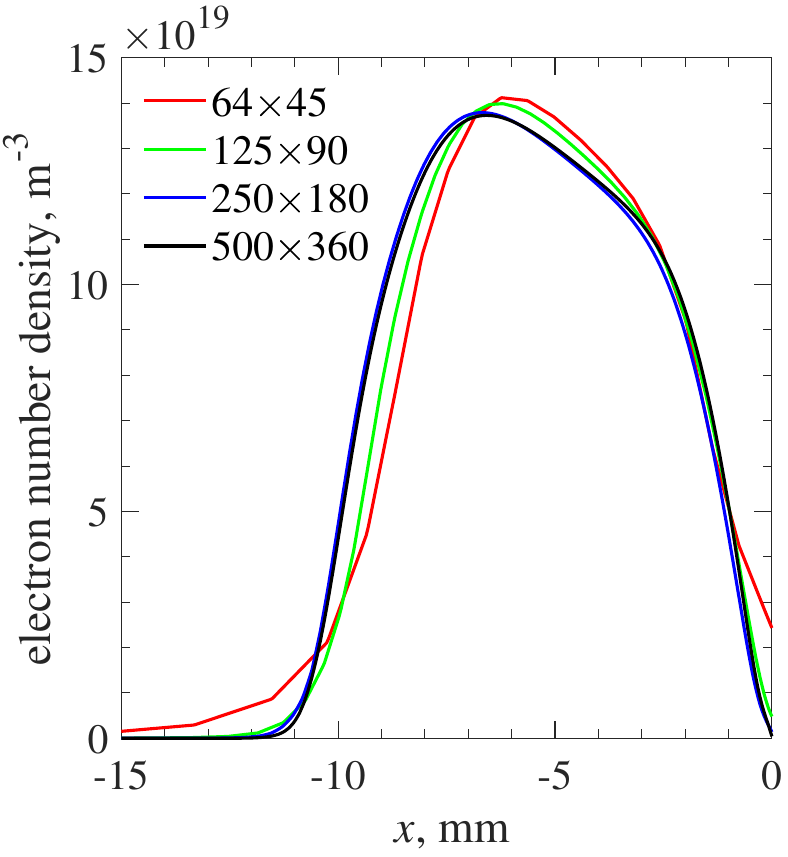}~~~}
     \subfigure[]{~~~\includegraphics[width=0.32\textwidth]{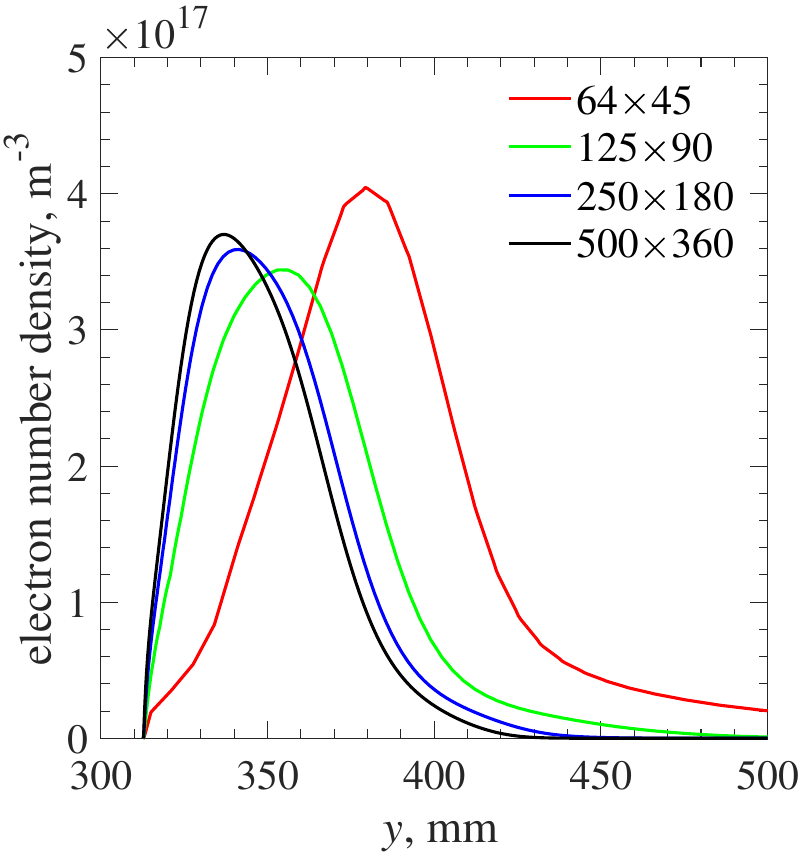}~~~}
     \figurecaption{Effect of grid size on (a) electron number density along the stagnation streamline and (b) electron number density 1 meter aft from the leading edge, for the 61~km altitude case.}
     \label{fig:RAMCII_grid_convergence_Ne}
\end{figure*}

\begin{figure*}[!h]
     \centering
     \subfigure[61 km altitude]{~~~\includegraphics[width=0.33\textwidth]{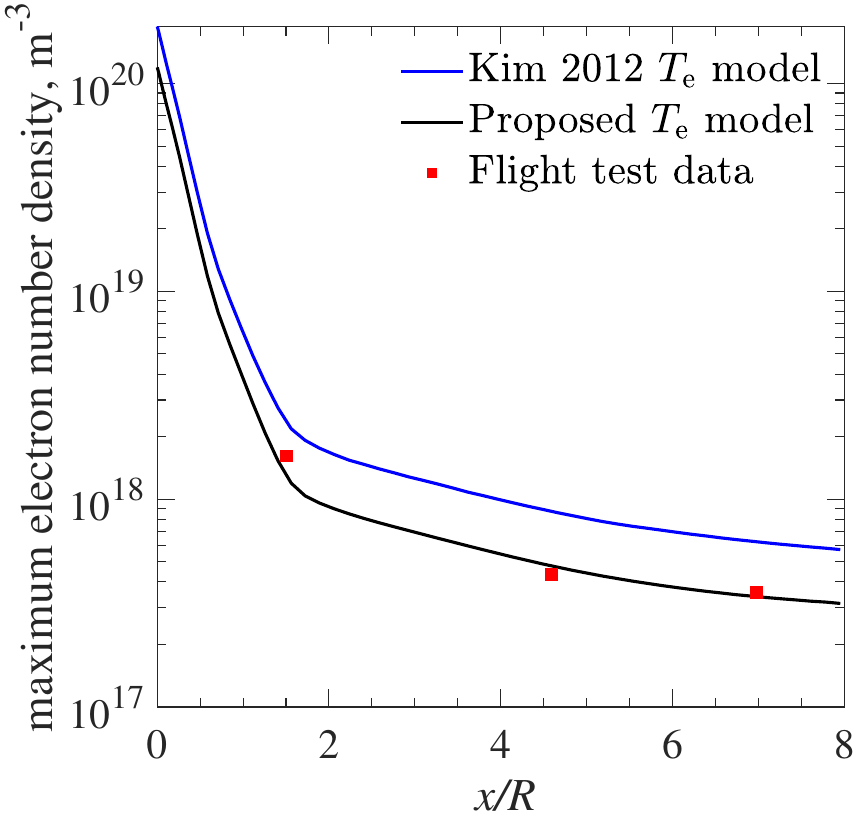}~~~}
     \subfigure[71 km altitude]{~~~\includegraphics[width=0.33\textwidth]{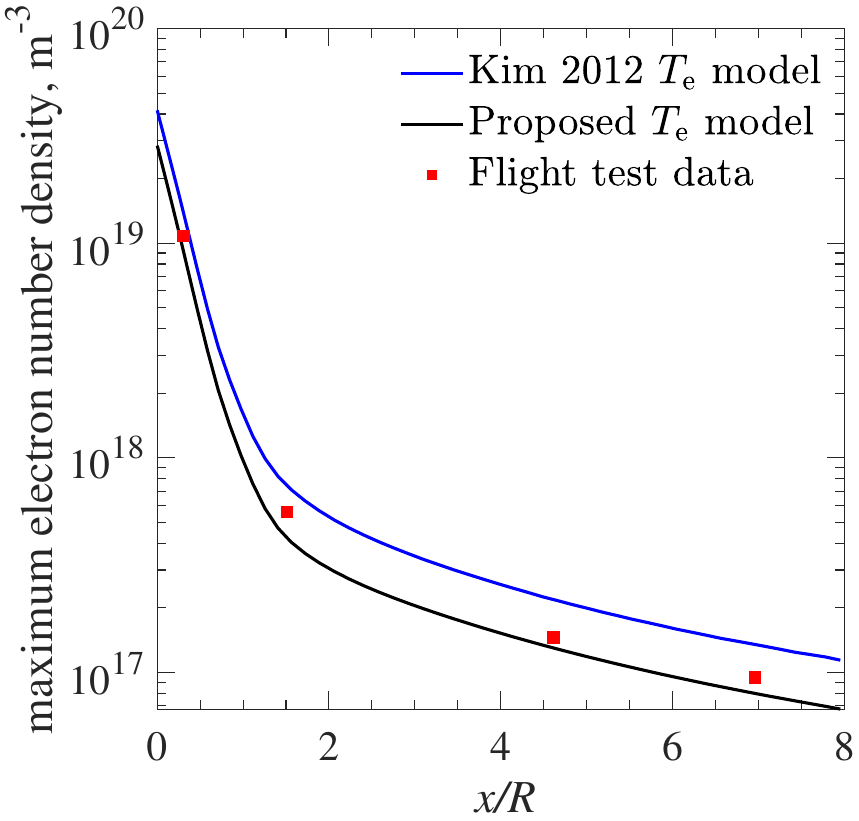}~~~}
     \figurecaption{Comparison between numerical results and RAM-C-II flight
test data on the basis of maximum electron number density along the vehicle axis at (a) 61
km altitude and (b) 71 km altitude.}
     \label{fig:RAMCII_Ne}
\end{figure*}
\begin{figure*}[!h]
     \centering
     \subfigure[Proposed model, 61 km]{~~~~\includegraphics[width=0.3\textwidth]{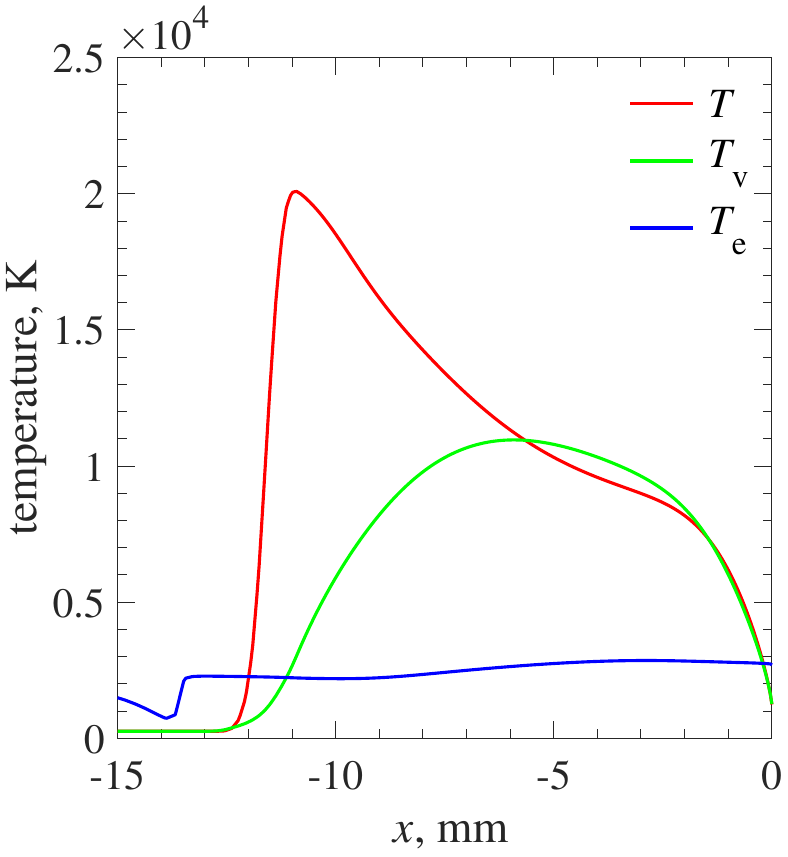}~~~~}
     \subfigure[Proposed model, 71 km]{~~~~\includegraphics[width=0.3\textwidth]{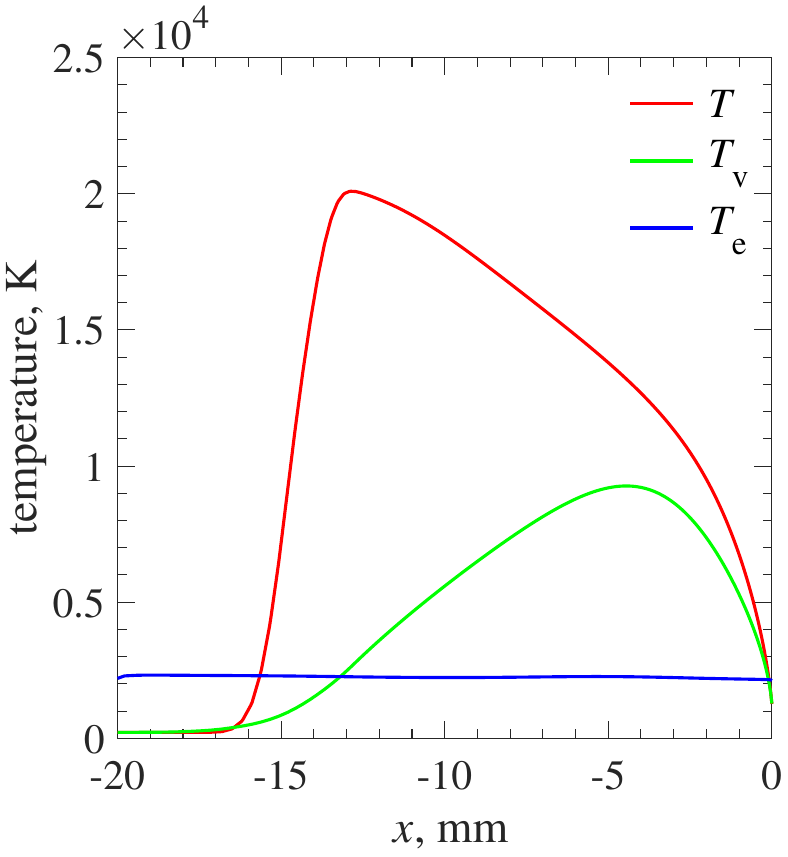}~~~~}\\
     \subfigure[Kim et al. model, 61 km]{~~~~\includegraphics[width=0.3\textwidth]{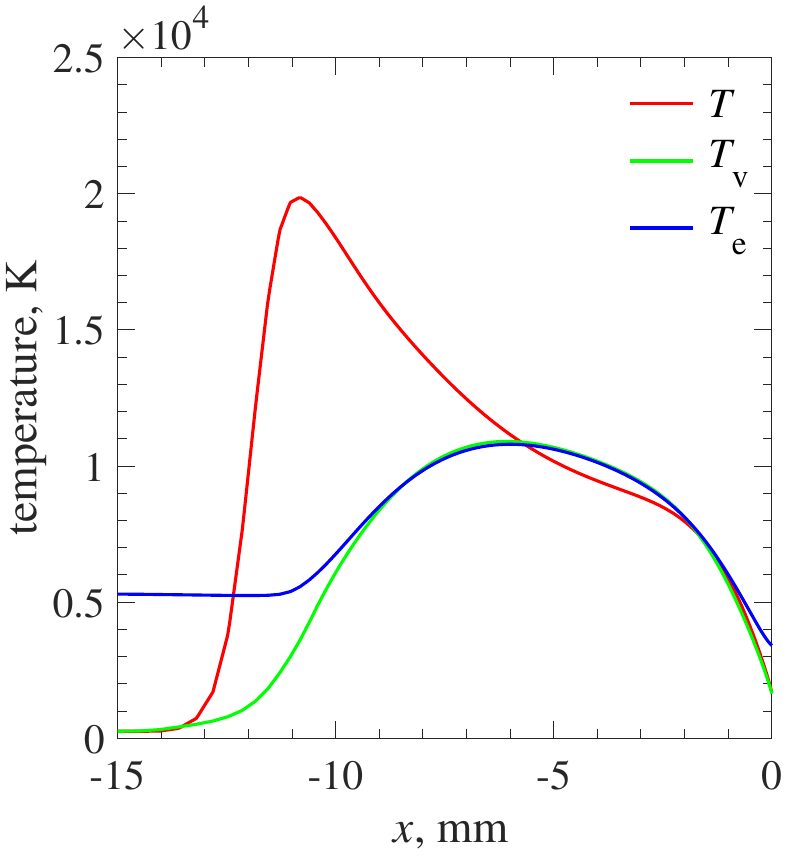}~~~~}
     \subfigure[Kim et al. model, 71 km]{~~~~\includegraphics[width=0.3\textwidth]{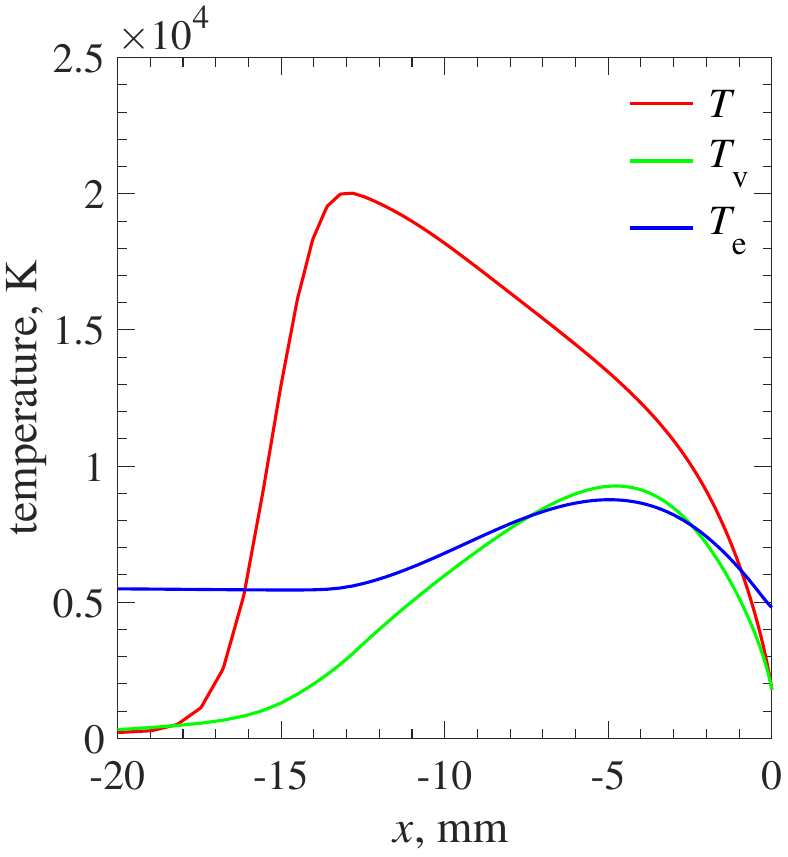}~~~~}
     \figurecaption{Comparison of translational, vibrational and electron temperature along the stagnation streamline for (a) the proposed model at 61~km, (b) the proposed model at 71~km, (c) the Kim et al. model at 61~km, and (d) the Kim et al. model at 71~km.}
     \label{fig:RAMCII_T_Tv_Te}
\end{figure*}
\begin{figure}[!h]
     \centering
     \includegraphics[width=0.47\textwidth]{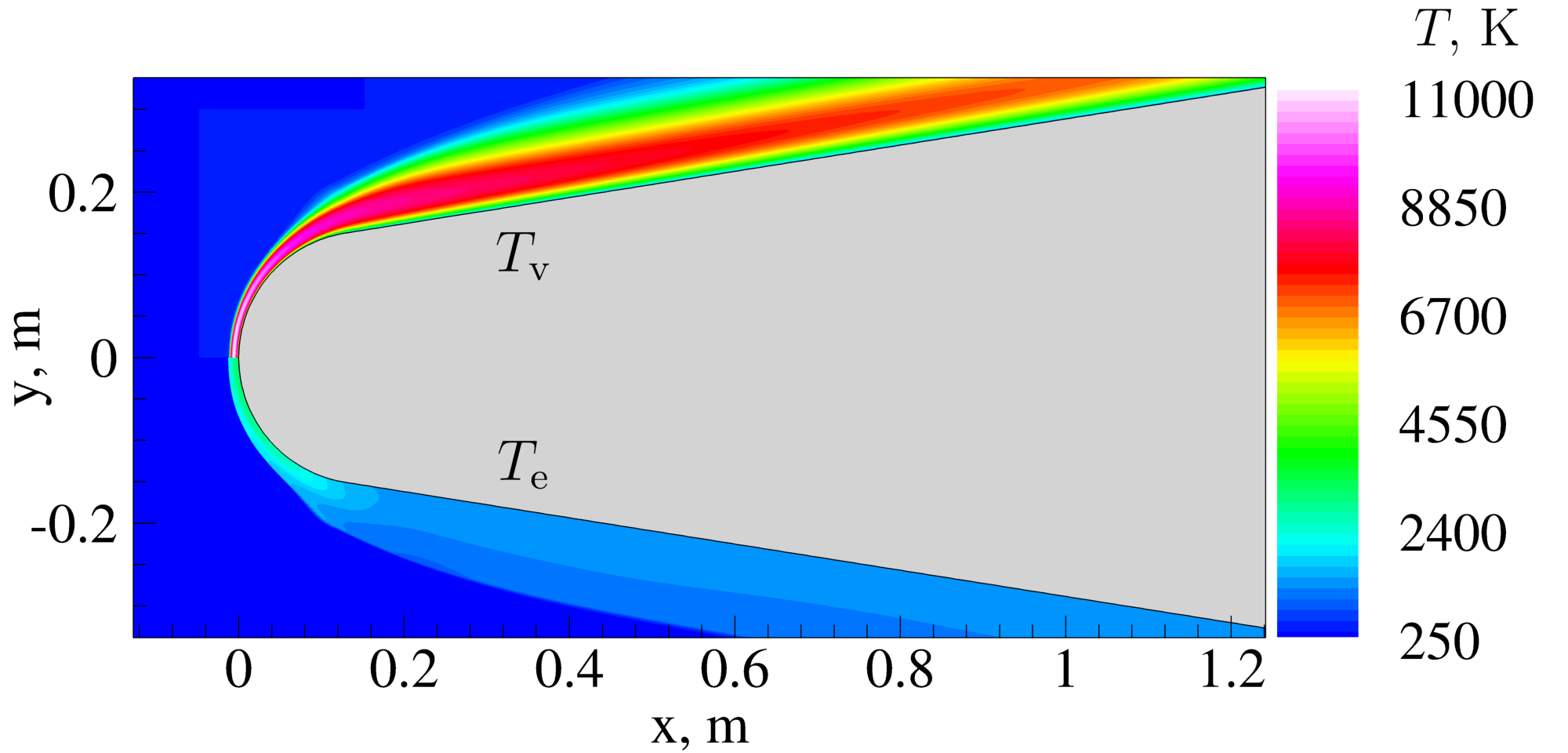}
     \figurecaption{Vibrational and electron temperatures around the RAM-C-II body for the 61 km altitude case obtained with the proposed model.}
     \label{fig:RAMCII_Te_Tv_contour}
\end{figure}

The second test case examined is an Earth entry flight test from the early 1970s known as RAM-C-II. This flight test aimed to measure the electron density in an axisymmetric plasma flow around a 1.2-meter long body with a blunt leading-edge radius of 15~cm, followed by a 15-degree truncated cone. Using microwave reflectometers, electron density was measured  at various points along the body  and at different altitudes.  At 61~km altitude, the Mach number was 23.9, and the freestream pressure was 19.962 Pa. At 71~km altitude, the Mach number was 25.9, and the freestream pressure was 4.844~Pa. For both altitudes, the velocity of the freestream with respect to the body was approximately 7650~m/s. 

For this test case, we employ a recently published high-temperature air 11-species chemical solver by \cite{ijhmt:2021:kim}, but with all reactions involving electrons in the reactants being assigned the rates shown in Table \ref{tab:correctedreactionrates}. Additionally, due to the high temperatures involved, we utilize the \cite{nasa:1990:gupta} model  to determine the mobilities and mass diffusion coefficients of all species, as well as the mixture viscosity and thermal conductivity. However, we adjust the electron mobility and electron thermal conductivity according to \cite{jtht:2023:parent} for better agreement with experimental data on the electrical conductivity and thermal conductivity of high-temperature air. No symmetry condition is used 
on the stagnation line because such leads to issues with the convergence of the potential equation used to find the electric field. The grid is clustered at the surface of the body to capture the non-neutral plasma sheath. For the baseline $125 \times 90$ mesh, the grid spacing at the surfaces varies between 10 micrometers (at the stagnation point) and 70 micrometers (at the trailing edge). As the mesh is refined, the grid spacing at the surfaces is reduced inversely proportional to the number of grid lines in each dimension. A grid convergence study of the electron density is presented in Fig.~\ref{fig:RAMCII_grid_convergence_Ne} for the 61~km case. Using a mesh of $500\times 360$ cells is seen to result in a numerical error on the maximum electron density of less than 5\%  on the stagnation line, and significantly less at other locations. All results shown hereafter are obtained with the $500\times360$ mesh.

In Fig.~\ref{fig:RAMCII_Ne}, a comparison is made between numerical results and experimental data based on the maximum electron number density within the boundary layer as a function of normalized axial distance (here, the axial distance is normalized with the leading-edge radius $R$). Excellent agreement (typically less than 5\% error) is achieved between the flight test data and the proposed electron energy model. This excellent agreement is observed at various locations along the body, as well as at altitudes of 61 km and 71 km.

In the same figure, we include a second set of numerical results obtained using the electron energy relaxation terms from the Kim 2012 model. For a fair comparison with the proposed electron energy source terms, only the electron energy relaxation source terms are varied, while the transport equations, chemical reactions, and transport coefficients remain unchanged. The Kim 2012 model is observed to yield electron number densities typically twice as large as the experimental values at an altitude of 61 km. At higher altitudes, the error is also significant, with a 30-40\% discrepancy observed with flight test data over most of the body.

\begin{figure}[!h]
     \centering
     \includegraphics[width=0.45\textwidth]{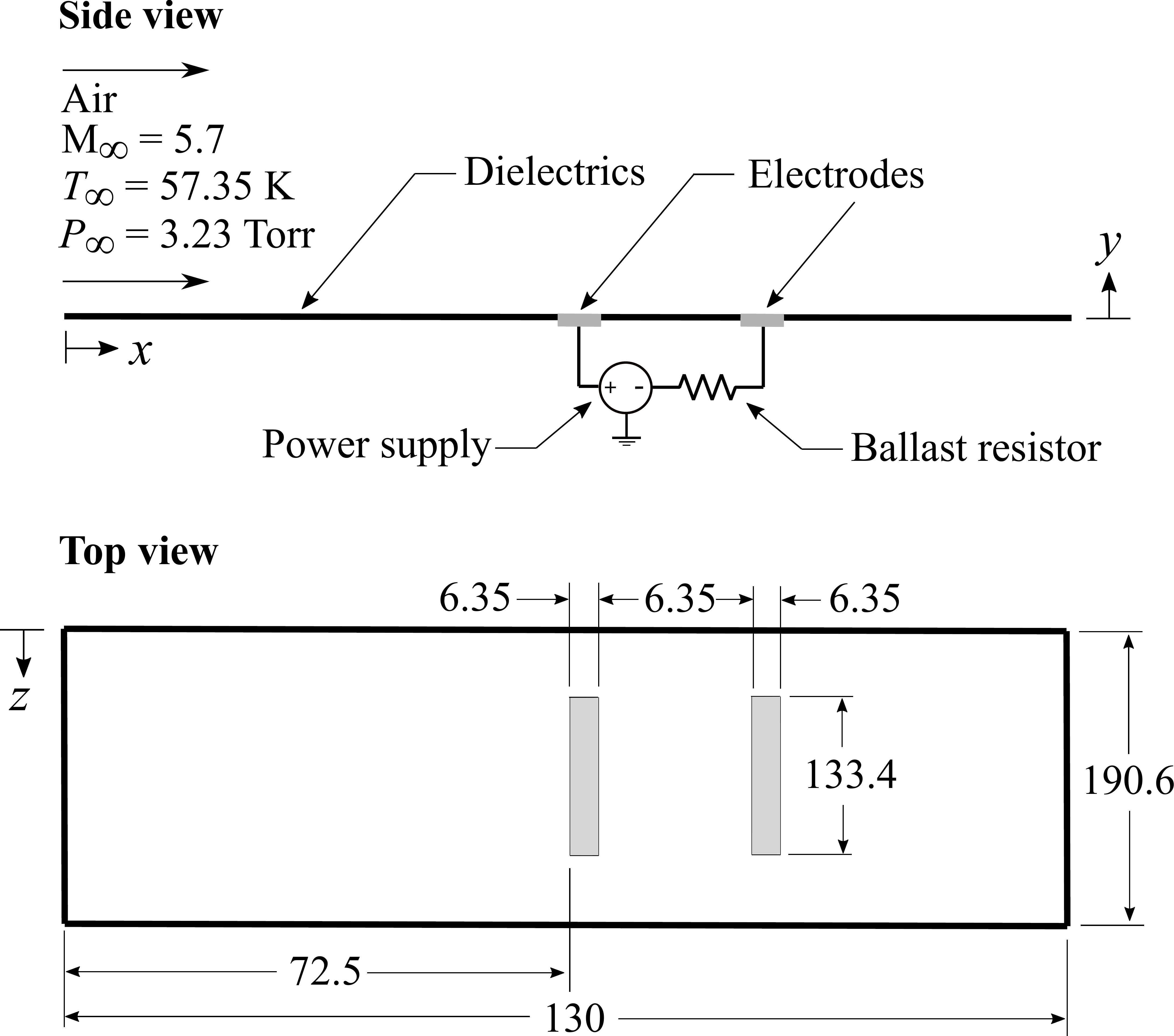}
     \figurecaption{Problem setup for the glow discharge  test case; all dimensions in mm.}
     \label{fig:discharge_setup}
\end{figure}

\begin{figure*}[!h]
     \centering
     \subfigure[plasma voltage, V]{~~~~\includegraphics[width=0.31\textwidth]{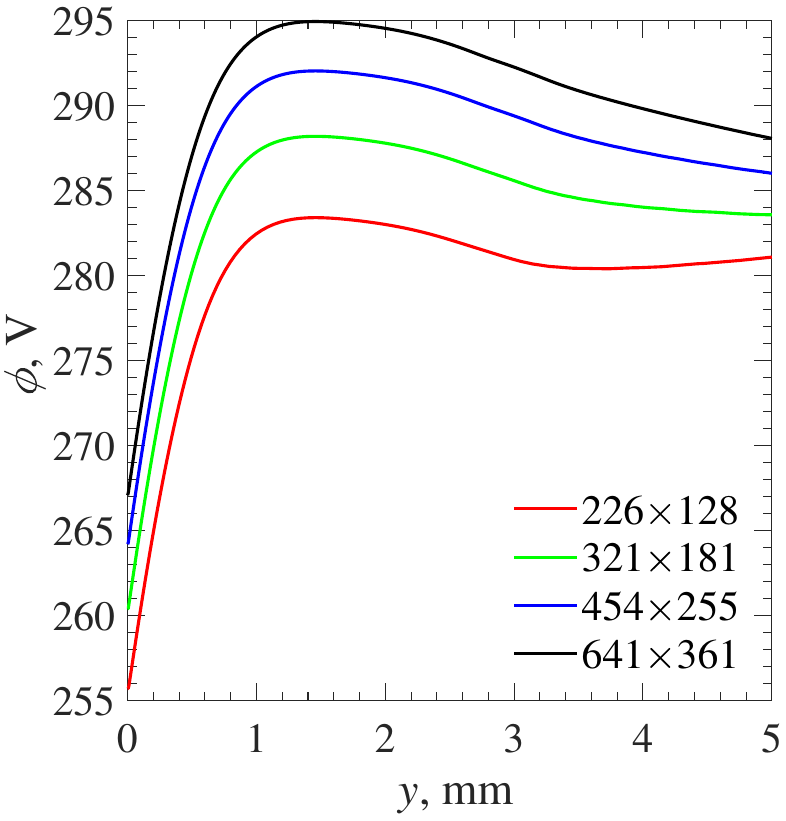}~~~~}
     \subfigure[electron number density, $\rm 1/m^3$]{~~~~\includegraphics[width=0.31\textwidth]{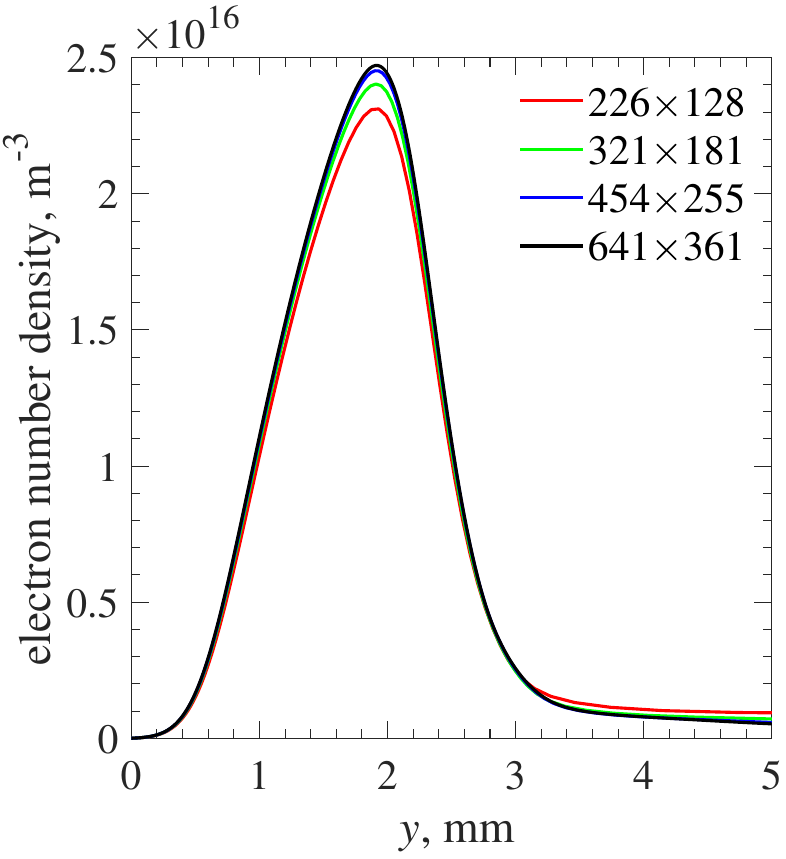}~~~~}
     \figurecaption{ Effect of grid size on (a) plasma voltage and (b) electron density  at the station $x=85$~mm using a 3000~V supply, a wall temperature of 350~K, and no ion mobility corrections.}
     \label{fig:discharge_grid_convergence}
\end{figure*}

\begin{figure*}[!h]
     \centering
     \subfigure[Uncorrected ion mobilities]{~~~~\includegraphics[width=0.38\textwidth]{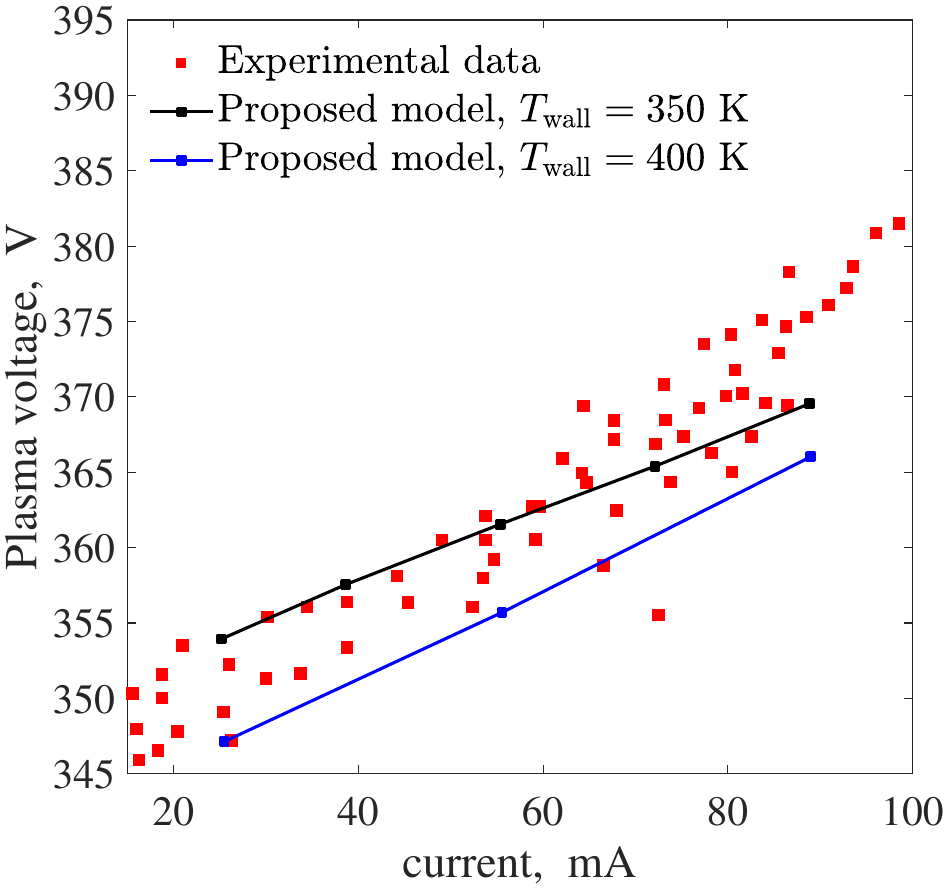}~~~~}
     \subfigure[Corrected ion mobilities]{~~~~\includegraphics[width=0.38\textwidth]{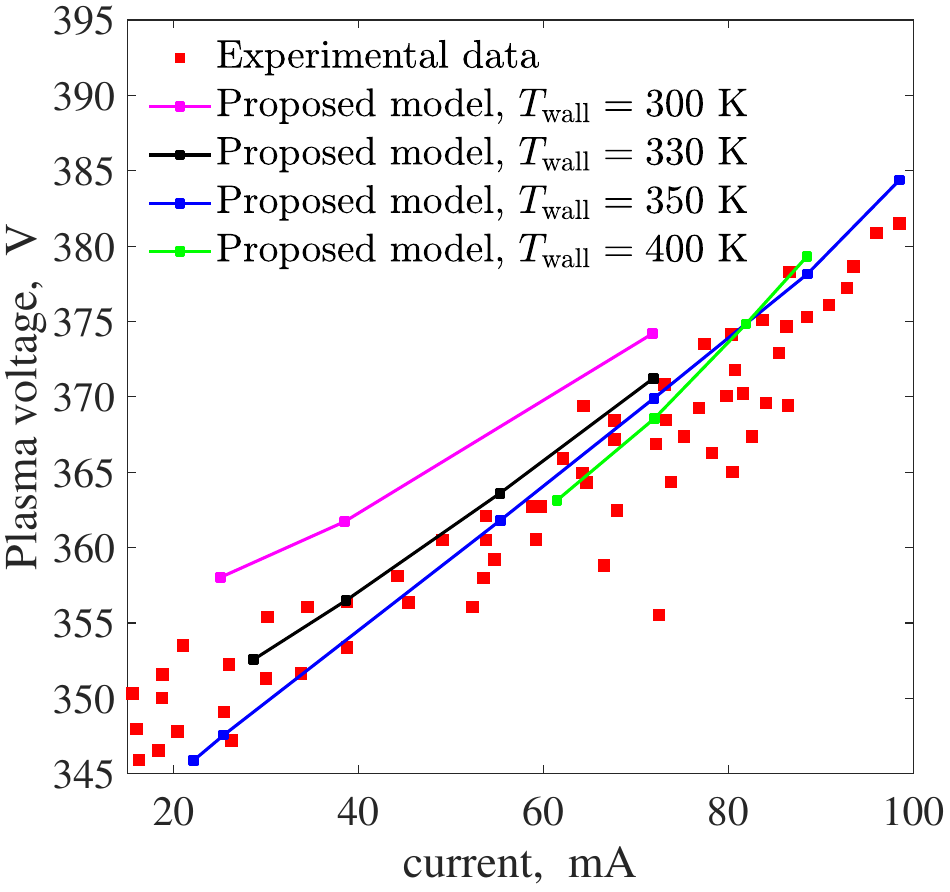}~~~~}
     \figurecaption{Comparison of proposed model  to experimental data on the basis of plasma voltage using (a) uncorrected ion mobilities and (b) corrected ion mobilities.}
     \label{fig:discharge_plasmavoltage_vs_current}
\end{figure*}

\begin{figure*}[!h]
     \centering
     \subfigure[electric field potential, V]{\includegraphics[width=0.38\textwidth]{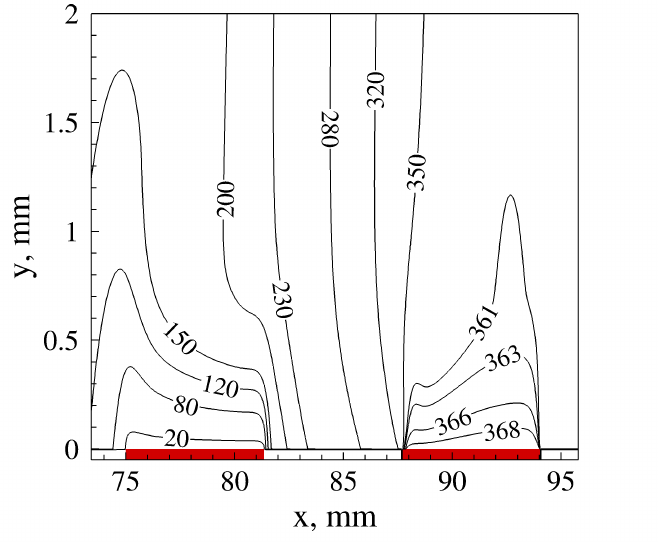}}
     \subfigure[electron temperature, eV]{\includegraphics[width=0.38\textwidth]{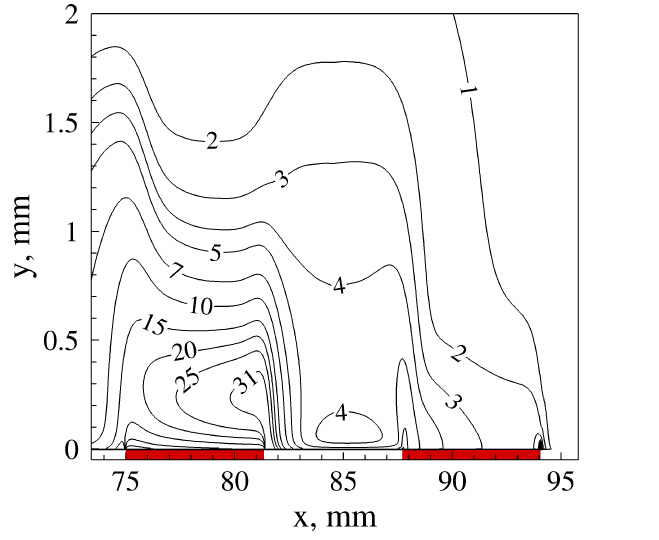}}
     \subfigure[electron number density, $10^{15}~\rm m^{-3}$]{\includegraphics[width=0.38\textwidth]{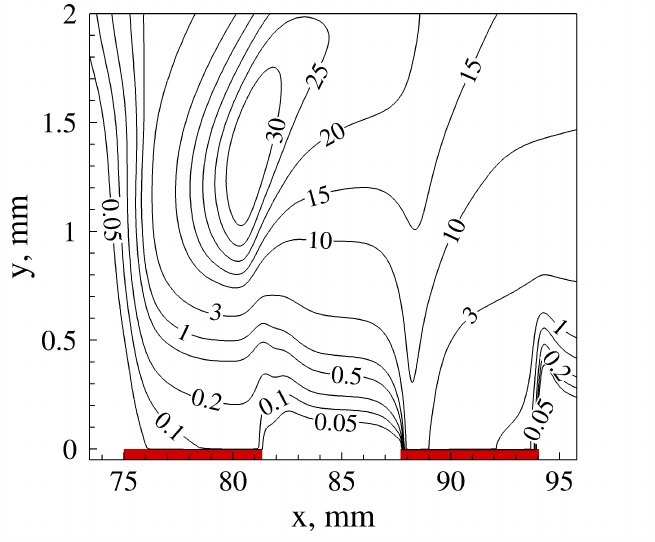}}
     \subfigure[electrical conductivity, $10^{-3}~\rm S/m$]{\includegraphics[width=0.38\textwidth]{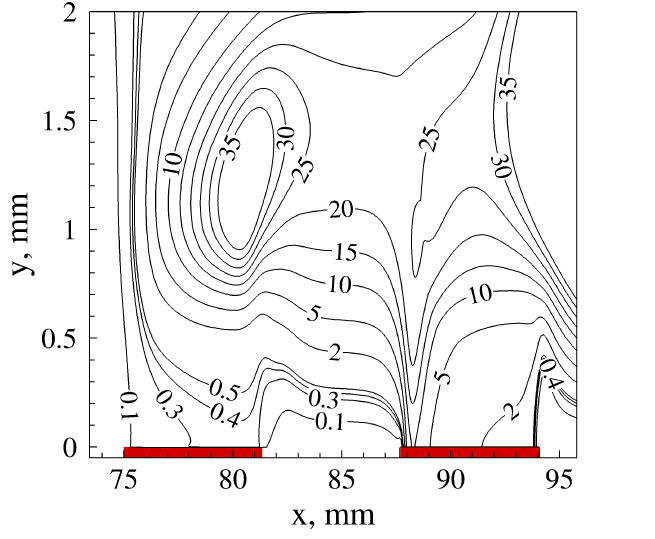}}
     \figurecaption{Contour levels of (a) $\phi$ in V, (b) $T_{\rm e}$ in eV, (c) $N_{\rm e}$ in $10^{15}$~m$^{-3}$, (d) electrical conductivity in $10^{-3}$~S/m, using a power source voltage of 3~kV,  a wall temperature of 350~K, and no correction to the ion mobilities.}
     \label{fig:discharge_contours}
\end{figure*}

Why does the Kim 2012 electron energy model result in a significantly higher electron number density than the proposed model? Since the electron temperature is determined from the electron energy transport equation, any alteration in the electron energy relaxation source terms directly influences the electron temperature.   A large difference in electron temperature can be observed downstream of the shock where the plasma density is the highest. Indeed, as can be observed from Fig.~\ref{fig:RAMCII_T_Tv_Te}, the electron temperature obtained with the Kim 2012 model is 4-5 times higher than the one obtained with the proposed model. This is due to the Kim 2012 model exhibiting much stronger relaxation terms in the temperature range 5,000~K -- 15,000~K. In turn, this leads to the electron temperature relaxing very quickly towards the vibrational temperature shortly downstream of the shock.  This does not occur with the proposed model which exhibits a large difference between vibrational and electron temperatures throughout the domain even at lower altitudes (see Fig.~\ref{fig:RAMCII_Te_Tv_contour}). Such a large difference in electron temperature between the Kim 2012 and the proposed model does not affect considerably electron gains because the electron temperature is not sufficiently high for Townsend ionization to be significant. However, it does lead to much reduced electron losses.  The primary electron loss mechanism in this scenario is 2-body dissociative recombination, with the rate scaling inversely proportional to the square root of the electron temperature. Therefore, given that the Kim 2012 model leads to a fourfold increase in electron temperature, it should correspondingly yield a twofold decrease in electron losses through 2-body dissociative recombination, and consequently, a twofold increase in electron density. This is a likely explanation for why the plasma density predicted by the Kim 2012 model is approximately double the experimental results.

Similar to the previous test case, we once again observe that the proposed electron energy model achieves excellent agreement with experimental data. Additionally, we again observe a significant disparity with the Kim 2012 model, which appears to be attributable to the latter over-predicting the rates of electron energy relaxation with the vibrational energy modes of the nitrogen molecule by several times.

\subsection{Glow Discharge in Hypersonic Boundary Layer}

The third test case involves a direct-current (DC) glow discharge acting on a hypersonic boundary layer, as recently investigated experimentally by  \cite{thesis:2022:broslawski}. The air inflow conditions and problem setup are specified in Fig.~\ref{fig:discharge_setup}, with the exception that the problem is simulated in 2D. Given that the boundary layer thickness and sheath thickness are orders of magnitude less than the depth of the domain, 3D effects near the center of the domain are expected to be negligible. Further, the density of the plasma is here high enough and the dimensions large enough that a fluid model for the charged species can be used with high accuracy. The fluid model validity  can be assessed through the relaxation distance associated with electron-neutral collisions. Such corresponds to the electron velocity divided by the electron-neutral collision frequency,   $L_{\rm relax}\sim V_{\rm e}/\nu_{\rm en}$. Starting from Eq.~(\ref{eqn:collisionfrequency}), we can express the collision frequency as a function of the reduced mobility as $\nu_{\rm en}=|C_{\rm e}| N /(m_{\rm e} \mu_{\rm e}^\star)$. We also can  approximate the electron velocity here as $V_{\rm e}\sim \mu_{\rm e} E$. After substituting the latter in the expression for the relaxation distance and reformatting, we obtain:
\begin{equation}
L_{\rm relax} \sim \frac{m_{\rm e}}{|C_{\rm e}| N}  E^\star (\mu_{\rm e}^\star)^2
\end{equation}
Knowing that the mixture density $N$ varies between $10^{23}$~m$^{-3}$ near the wall and $5\cdot 10^{23}$~m$^{-3}$ in the freestream, and using Figs.~\ref{fig:Estar_Te_splines} and \ref{fig:mueN_Te_splines} to find the reduced electric field and reduced mobility, the relaxation distance is found to vary between 0.2 and 40 micrometers in the electron temperature range 1-30 eV. As will be shown subsequently, such relaxation distances are shorter than the distances for which the electron temperature changes significantly within the discharge, thus confirming the validity of the fluid model for this problem.

The wall temperature was observed to vary between 300~K and 450~K while performing the experiments and to vary not only along the plate but also from shot to shot. Because the temperature distribution is not known, we here fix the wall temperature  over the whole flat plate to a constant which we will either set to 300, 330, 350, or 400~K. The air density is sufficiently low that the DC discharge remains filament-free and steady. Due to the low temperature of the air, we utilize the \cite{book:1984:dixon-lewis} transport coefficients, along with the electron mobilities specified in Table~\ref{tab:mueNfromTe} and ion mobilities taken from \cite[Table 8.1]{book:2022:parent} with and without the high electric field correction. Additionally, we employ a low-temperature 8-species chemical model outlined in \cite{jpp:2007:parent}, adjusted with the electron reactions outlined in Table~\ref{tab:correctedreactionrates}.

 We can estimate the secondary electron emission (SEE) coefficient on the cathode using the correlation outlined in \cite[pages 56, 134, 180]{book:1991:raizer}. First, we note that the largest experimentally-measured current is about 80~mA, resulting in the largest average current density on the cathode being 95 A/m$^2$. Additionally, the gas temperature and pressure near the cathode are known to be 350~K and 3.2~Torr. Using these values, and for a specified SEE coefficient of 0.1--0.3, Raizer's correlation predicts a voltage drop across the cathode sheath in the range of 0.4--1.6 kV. Since this exceeds the maximum measured plasma voltage of 375~V, it implies that the SEE coefficient should be higher than 0.3. Consequently, we set the SEE coefficient to a higher value of 0.6. With this adjustment, Raizer's correlation predicts a cathode sheath voltage drop of 189 V. This seems to be a more accurate estimate for the SEE coefficient, as it results in the cathode sheath voltage being about half of the plasma voltage. Further, this aligns well with experimental data of effective electron yield per ion in \cite[Fig.~5]{psst:1999:phelps} for a similar cathode reduced electric field as the one predicted by the Raizer correlation for this discharge  (approximately 20,000~Td).  It is  noted that the Raizer correlation is used here with the same ion mobility (excluding electric field effects) as the one employed in the Dixon-Lewis transport model.  When the effect of the electric field on ion mobility is included, we find that the SEE coefficient needs to be set to an even higher value around 0.8 for the cathode sheath voltage drop to be approximately half of the plasma voltage. Therefore, we here set on the electrodes $\gamma_{\rm e} =0.6$  when no electric field correction is applied to the ion mobilities and $\gamma_{\rm e} =0.8$  when the electric field correction is applied to the ion mobilities.

Being 2D, steady, and strongly dependent on electron temperature, this is an ideal case to validate our proposed electron energy source terms, as the grid-induced error can be relatively well estimated and the numerical error minimized. The mesh is constructed so that the number of grid lines in each dimension scales with the mesh factor and, consequently, that the spacing between cells scales inversely proportional to the mesh factor. For a mesh factor of 1, the grid consists of $161 \times 91$ cells, with clustering near the wall to ensure a spacing of 10 micrometers between the wall and the near-wall node. In Fig.~\ref{fig:discharge_grid_convergence}, the effect of the grid (obtained with a mesh factor varying between $\sqrt{2}$, 2, $2\sqrt{2}$, and 4) on the electric field potential and the electron number density is shown for a ballast resistance of 30 k$\Omega$, a power supply voltage of 3~kV, a SEE of 0.6 and no electric field correction applied to the ion mobilities.  Using Richardson extrapolation, we estimate that a grid with a mesh factor of $2\sqrt{2}$ leads to at most a 3.3\% error on the voltage and at most a 1\% error on the electron density. This grid design comprised of $454 \times 255$ cells and a near wall node spacing of 3.5 microns is subsequently used to obtain results.

Some properties measured with high accuracy during the experiments include the current flowing from one electrode to the other as a function of plasma voltage and power source voltage as depicted in \cite[Fig.~7.8]{thesis:2022:broslawski}. Here, we focus on the plasma voltage (i.e., the voltage difference across electrodes) rather than the power supply voltage because the latter is not very sensitive to what occurs in the plasma. Indeed, because the ballast resistance of 30~k$\Omega$ is higher than the plasma resistance, most of the power supply voltage is lost to heat deposited within the ballast, not in the plasma. In Fig.~\ref{fig:discharge_plasmavoltage_vs_current}, a comparison between experimental and computational results is shown on the basis of plasma voltage as a function of current.  The voltage-current relationship is here challenging to capture accurately because the amount of current flowing between electrodes is very sensitive to the plasma voltage:  the plasma voltage varies by only 5\% (between 350 and 370 V) for a current varying by more than 4 times (between 20 and 90 mA). Nonetheless, very good agreement can be observed in this regard with a very similar trend observed between the proposed model and the experiments especially when applying the electric field correction to the ion mobilities.

Obtaining very good agreement for this test case is an indicator that the proposed electron temperature source terms are accurate. Indeed, given a certain plasma voltage, the current is proportional to the electrical conductivity, which is proportional to the product of electron density and electron mobility. Both electron density and electron mobility are highly dependent on electron temperature. Electron mobility varies by more than one order of magnitude for the range of electron temperature experienced here (see Fig.~\ref{fig:mueN_Te_splines} and Fig.~\ref{fig:discharge_contours}b). Furthermore, the electron density shown in Fig.~\ref{fig:discharge_contours}c is a balance between electron creation through Townsend ionization and electron destruction through 2-body dissociative recombination. Both processes are dependent on electron temperature (see Fig.~\ref{fig:reactionrate_Te_splines}). Moreover, not only does this problem test the validity of the electron energy modeling, but it does so over a large range of electron temperatures. As can be seen from Figs.~\ref{fig:discharge_contours}a and \ref{fig:discharge_contours}b, about half the voltage is lost in the positive column and the anode sheath where the electron temperature is in the range 3-7 eV, while the other half is lost in the cathode sheath where the electron temperature is in the range 7-30 eV. The very good agreement on the basis of plasma voltage versus current observed here between experiments and numerical results is an indicator that the electron energy source terms proposed herein are accurate in both of these ranges.

\section{Conclusions}

A novel formulation of the electron energy relaxation terms is presented here, applicable to non-neutral and quasi-neutral plasma flows where the electron temperature could be higher or lower than the gas temperature. The present approach is advantageous over previous models for plasma flows by expressing all inelastic electron energy losses for each species in terms of the reduced electric field and the reduced electron mobility. This contrasts with prior models which required separate cross-sectional data for each electron energy loss mechanism, such as Townsend ionization, vibrational excitation, rotational excitation, etc.  

The approach proposed herein leads to a more accurate modeling of electron energy relaxation when the reduced electron mobility and the reduced electric field can be readily obtained as a function of electron temperature from experiments. It is emphasized that this fully accounts for all electron energy losses due to electron impact processes (such as vibrational excitation, Townsend ionization, etc.) without the need to quantify each process independently. This is particularly advantageous for molecular gases because the latter have more electron excitation processes (vibrational and rotational), which are difficult to measure accurately over a wide range of energies, apart from other processes. 

The new model is validated through several test cases, including plasma flows where the electron temperature is lower than the gas temperature, as well as discharges where the electron temperature exceeds 30 eV. In all cases, very good to excellent agreement with experimental data is observed, significantly surpassing prior electron energy models for plasma flows.

\section*{Acknowledgments}

This study has been sponsored in part by Tokyo Electron Ltd. The Authors would like to thank Mikhail Shneider, \textcolor{black}{Sung-Min Jo}, Sergey Macheret, Jozef Brcka, Raymond Joe, Anthony Dip, and \textcolor{black}{Naoshige Fushimi}  for fruitful discussions that helped improve this manuscript.


\footnotesize
\bibliography{all}
\bibliographystyle{plainnatmod}

\end{document}

\begin{figure}[ht!]
     \centering
     \subfigure[(a) blah blah]{\includegraphics[width=0.49\textwidth]{figs/fig1.pdf}}
     \subfigure[(b) blah blah]{\includegraphics[width=0.49\textwidth]{figs/fig2.pdf}}
     \figurecaption{Caption here.}
     \label{fig:name}
\end{figure}

\begin{table}[t]
  \center\fontsizetable
  \begin{threeparttable}
    \tablecaption{Additional Townsend ionization and recombination rates for N$_2$, O$_2$, N, O, and NO.\tnote{a}}
    \label{tab:townsend}
    \fontsizetable
    \begin{tabular*}{\textwidth}{l@{\extracolsep{\fill}}lll}
    \toprule
    No.&Reaction & Rate Coefficient  & Refs. \\
    \midrule
    29  & $\rm e^- + N_2   \rightarrow N_2^+ + e^- + e^-$  
       &  ${\rm exp}(-0.0105809\cdot {\rm ln}^2 E^\star - 2.40411\cdot 10^{-75} \cdot {\rm ln}^{46}E^\star)$~cm$^3$/s
       & \cite{jcp:2014:parent} \\
    30  & $\rm e^- + O_2   \rightarrow O_2^+ + e^- + e^-$  
       &  ${\rm exp}(-0.0102785\cdot {\rm ln}^2 E^\star - 2.42260\cdot 10^{-75} \cdot {\rm ln}^{46}E^\star)$~cm$^3$/s
       & \cite{jcp:2014:parent} \\
    \bottomrule
    \end{tabular*}
\begin{tablenotes}
\item[{a}] Notation and units: ${\cal A}$ is Avogadro's number set to $6.02214 \times 10^{23}$ mol$^{-1}$; $E^\star$ is the reduced effective electric field ($E^\star\equiv|\vec{E}|/N$) in units of V$\cdot$m$^2$.

\end{tablenotes}
   \end{threeparttable}
\end{table}